\documentclass{aastex63}

\hypersetup{linkcolor=red,citecolor=blue,filecolor=cyan,urlcolor=magenta}

\submitjournal{ApJ}

\shorttitle{Companion to M13F}
\shortauthors{M. Cadelano et al.}
\graphicspath{{./}{figures/}}
\usepackage{mathrsfs}  
\usepackage{color,graphicx}
\begin{document}

\title{PSR J1641$+$3627F: a low-mass He white dwarf orbiting a possible high-mass neutron star \\ in the globular cluster M13}

\correspondingauthor{Mario Cadelano}
\email{mario.cadelano@unibo.it}

\author[0000-0002-5038-3914]{Mario Cadelano}
\affil{Dipartimento di Fisica e Astronomia, Università di Bologna, Via Gobetti 93/2 I-40129 Bologna, Italy}
\affil{INAF-Osservatorio di Astrofisica e Scienze dello Spazio di Bologna, Via Gobetti 93/3 I-40129 Bologna, Italy}

\author[0000-0000-0000-0000]{Jianxing Chen}
\affil{Dipartimento di Fisica e Astronomia, Università di Bologna, Via Gobetti 93/2 I-40129 Bologna, Italy}
\affil{INAF-Osservatorio di Astrofisica e Scienze dello Spazio di Bologna, Via Gobetti 93/3 I-40129 Bologna, Italy}

\author[0000-0002-7104-2107]{Cristina Pallanca}
\affil{Dipartimento di Fisica e Astronomia, Università di Bologna, Via Gobetti 93/2 I-40129 Bologna, Italy}
\affil{INAF-Osservatorio di Astrofisica e Scienze dello Spazio di Bologna, Via Gobetti 93/3 I-40129 Bologna, Italy}

\author[0000-0002-8811-8171]{Alina G. Istrate}
\affil{Department of Astrophysics/IMAPP, Radboud University Nijmegen, PO Box 9010, NL-6500 GL Nijmegen, the Netherlands}

\author[0000-0002-2165-8528]{Francesco R. Ferraro}
\affil{Dipartimento di Fisica e Astronomia, Università di Bologna, Via Gobetti 93/2 I-40129 Bologna, Italy}
\affil{INAF-Osservatorio di Astrofisica e Scienze dello Spazio di Bologna, Via Gobetti 93/3 I-40129 Bologna, Italy}

\author[0000-0001-5613-4938]{Barbara Lanzoni}
\affil{Dipartimento di Fisica e Astronomia, Università di Bologna, Via Gobetti 93/2 I-40129 Bologna, Italy}
\affil{INAF-Osservatorio di Astrofisica e Scienze dello Spazio di Bologna, Via Gobetti 93/3 I-40129 Bologna, Italy}

\author[0000-0003-1307-9435]{Paulo C. C. Freire}
\affiliation{Max-Planck-Institut für Radioastronomie MPIfR, Auf dem Hügel 69, D-53121 Bonn, Germany}

\author[0000-0002-2744-1928]{Maurizio Salaris}
\affil{Astrophysics Research Institute, Liverpool John Moores University, 146 Brownlow Hill, Liverpool L3 5RF, UK}

\begin{abstract}

We report on the discovery of the companion star to the millisecond pulsar J1631+3627F in the globular cluster M13. By means of a combination of optical and near-UV high-resolution observations obtained with the Hubble Space Telescope, we identified the counterpart at the radio source  position. Its location in the color-magnitude diagrams reveals that the companion star is a faint ($V\approx 24.3$) He-core white dwarf. We compared the  observed companion magnitudes with those predicted by state-of-the-art binary evolution models and found out that it has a mass of $0.23\pm0.03 \ M_{\odot}$, a radius of $0.033^{+0.004}_{-0.005} \ R_{\odot}$ and a surface temperature of $11500^{+1900}_{-1300}$ K. Combining the companion mass with the pulsar mass function is not enough to determine the orbital inclination and the neutron star mass; however, the last two quantities become correlated: we found that either the system is observed at a low inclination angle, or the neutron star is massive. In fact, assuming that  binaries are randomly aligned with respect to the observer line of sight, {there is a $\sim70\%$ of probability that this system hosts a neutron star more massive than $1.6 \ M_{\odot}$. In fact,} the maximum and median mass of the neutron star, corresponding to orbital inclination angles of $90^{\circ}$ and $60^{\circ}$, are $M_{NS,max}=3.1\pm0.6 \ M_{\odot}$ and $M_{NS,med}=2.4\pm0.5 \ M_{\odot}$, respectively. {On the other hand, assuming also an empirical neutron star mass probability distribution, we found that this system could host a neutron star with a  mass of $1.5\pm0.1 \ M_{\odot}$ if orbiting with a low-inclination angle around $40^{\circ}$.}


\end{abstract}


\section{Introduction} \label{sec:intro}

\subsection{Millisecond Pulsars}
Millisecond pulsars (MSPs) are rapidly spinning neutron stars (NSs) usually formed in binary systems through mass-transfer from an evolving low-mass companion star \citep[$\lesssim 1.4 \  M_{\odot}$;][]{alpar82,bhattacharya91,handbook}. Active mass-transfer is commonly observed in binaries with a NS primary, generally considered as the precursors of MSPs \citep[e.g.][]{tauris06,archibald10,papitto13,stappers14,ferraro15}. Following the mass-transfer phase, the NS is reactivated as a pulsar and it is usually observable in the radio bands, while the companion is expected to be the core-remnant of an exhausted and stripped evolved star: a white dwarf with a He core \citep[He-WD; e.g.][]{driebe98,tauris99,ferraro03,istrate14a,antoniadis16wd,cadelanotesi}.
Although this scenario has been firmly confirmed throughout the years, several deviations from it have been highlighted with the discovery, for example, of MSPs with massive CO-WD companions \citep[e.g.][]{tauris11,pallanca13wd}, double NS systems \citep[e.g.][]{jacoby06,tauris17,ridolfi19} or eclipsing MSPs with non-degenerate companion stars \citep[e.g.][]{pallanca10,breton13,mucciarelli13,roberts13,cadelano15_m71,roberts18}.

\subsection{Neutron Star Masses}

The bulk of the NS population has masses around $1.4 \ M_{\odot}$. However, recent studies on the NS mass function \citep{antoniadis16} found evidence of a bi-modal distribution peaked at $1.4 \ M_{\odot}$ and $1.8 \ M_{\odot}$ and suggested a limiting sustainable NS mass of $\geq 2.018 \ M_{\odot}$, before the stellar structure collapses to form a black-hole. NS mass measurements are driven by one of the most important and still unanswered question of modern physics: the behavior of cold matter at densities larger than that of nuclear saturation. Indeed, this regime is found in NS interiors, and constraining their still unknown equation of state therefore is one of the key ingredient to answer such a long-lasting unsolved question \citep[e.g.][]{lattimer01,lattimer07,steiner10,ozel16,bogdanov19}. The increasing number of NS mass measurements and in particular the discovery of very high-mass NSs \citep{demorest10,antoniadis13,fonseca16,cromartie20} already helped to place constraints on the NS equation of state, leaving however a vast range of possibilities. \citep[][and references therein]{ozel16,antoniadis16}. 

In the case of MSPs with degenerate companions, such as WDs or NSs, precise mass measurements of both the binary components can be directly obtained through the timing analysis if relativistic effects are observed \citep[e.g. periastron precession, Shapiro delay, etc...; see Section 2 in][]{ozel16}. If they are not, mass measurements can be indirectly performed through the analysis of the WD companions, usually observable through UV and optical observations. In fact, by comparing the UV-optical magnitudes with appropriate binary evolution models, the mass of the companion star can be evaluated and such a value, combined with the orbital properties of the system, can place constraints on the NS mass \citep[e.g.][]{cadelano15_47tuc,cadelano19,dai17,kirichenko20}. Furthermore, {spectroscopic observations can also be used to determine the companion radial velocity curve. This can be combined with the radial velocity curve of the NS to measure the binary mass ratio, thus providing additional constraints on the NS mass \citep[e.g.][]{ferraro03massratio}. In some cases, the spectroscopic data is good enough to allow the determination of the WD mass as well, which combined with the mass ratio allows the determination of the NS mass \citep{antoniadis12,antoniadis13,mata20}. Spectroscopic studies of WD companions have been also very useful for tests of gravity theories \citep[e.g.][]{antoniadis13}\footnote{An updated list of NS masses determined from timing and precise optical measurements can be found at \url{https://www3.mpifr-bonn.mpg.de/staff/pfreire/NS_masses.html}}.}

\subsection{Pulsars in globular clusters and their companions}

While MSPs are ubiquitous throughout the whole Galaxy, their formation rate is enhanced by a factor of $10^3$ in globular clusters
(GCs). In fact, GCs are collisional systems where internal dynamics promote the formation of exotic systems like blue straggler stars, cataclysmic variables and MSPs \citep{ferraro09,ferraro16ebss,hong17,cadelano18,riverasandoval18,campos18}, {the latter sometimes in eccentric orbits with massive companions acquired by exchange encounters after the NS was already recycled \citep[e.g.][and references therein]{ridolfi19}}. This makes GCs the ideal environment for studying exotic stellar systems, which can also be used as test particles in modelling the complex interplay between cluster internal dynamics and stellar evolution \citep{ferraro12,ferraro18,ferraro19,cheng19,cheng1947tuc}. Moreover, they can provide a wealth of information about the physical properties of the host cluster itself \citep[e.g.][]{prager17,freire17,abbate18,abbate19}. 

We are currently leading a long-term program aimed at identifying MSP companions in Galactic GCs. This program led to several discoveries and characterisation of He-WD companions as well as non-degenerate and more exotic systems \citep{ferraro01,ferraro03,cocozza06,pallanca10,pallanca13,pallanca14,mucciarelli13,cadelano15_47tuc,cadelano15_m71,cadelano17x,cadelano19,cadelanotesi}, shedding light on binary and stellar evolution under extreme conditions. In this work, we report on the discovery of the companion star to the recently discovered PSR J1641+3627F (hereafter M13F) in the GC M13 \citep{wang20}.

M13 (NGC 6205) is a low-density cluster located at about 7 kpc from the Sun \citep[][2010 edition]{harris96} and affected by a very low stellar extinction $E(B-V)=0.02$ \citep{ferraro99}. Its stellar population has an age around 13 Gyr \citep{dotter10}, an intermediate-low metallicity of $[Fe/H]=-1.5$ \citep{carretta09} and an extended blue horizontal branch \citep{ferraro97}. This cluster hosts six MSPs \citep{kulkarni91,hessels07,wang20}: two are isolated NSs, three are canonical systems likely having a WD companion and one is an eclipsing binary. M13F is one of the three canonical systems and was recently discovered with the FAST radiotelescope by \citet{wang20}. It has a spin period of 3 ms and an almost circular orbit of 1.4 days. The main properties of the system, useful for this work, are summarized in Table~\ref{tab:m13f}. The binary orbital parameters result in a NS mass function $f=0.001108878 \ M_{\odot}$. Such a value, assuming a typical NS mass of $1.4 \ M_{\odot}$, implies an extremely low minimum companion mass (corresponding to an edge-on orbit) of $0.13 \ M_{\odot}$ and a median companion mass (corresponding to an inclination angle of $60^{\circ}$) of $0.16 \ M_{\odot}$. However, state-of-the-art binary evolution models \citep[][]{istrate14a,istrate14b,istrate16} show that companion stars with intermediate-low metallicities (such that of stars in M13) are very unlikely to create {a detached He-core WD} less massive than $\sim0.18 \ M_{\odot}$ by the end of the mass-transfer and the binary detachment. This is mainly due to the fact that low-metallicity stars have shorter evolutionary timescales with respect to high-metallicity stars, and (having smaller radii) they are able to fill their Roche-Lobe at later stages of the binary evolution.
This suggests that M13F could be a MSP with a non canonical companion star, i.e., a star different from a He-WD. Alternatively, it could be either a nearly face-on system, or a binary containing a NS more massive than $1.4 \ M_{\odot}$ (in these cases, in fact, the companion mass would be larger than $0.18 \ M_{\odot}$). All this motivates an investigation about the true nature and mass of the companion to M13F.


The paper is outlined as follows: in Section~\ref{sec:data} the data-set and data reduction procedure are described; in Section~\ref{sec:comp} we present the identification of the companion star to M13F and we compare its properties with those predicted by binary evolution models. Finally, in Section~\ref{sec:conc} we draw our conclusions.

\begin{deluxetable*}{lcc}
\tablecolumns{3}
\tablewidth{0pt}
\tablecaption{Main radio timing
parameters for M13F\tablenotemark{a}, from \citet{wang20}.\label{tab:m13f}}
\tablehead{\colhead{Parameter} & \colhead{Value}}
\startdata
Right ascension, $\alpha$ (J2000)\dotfill  & 16$^{\rm h}\,41^{\rm m}\,44\fs6058(3)$     \\
Declination, $\delta$ (J2000)  \dotfill  & $36^\circ\,28'\,16\farcs0034(2)$     \\
Angular offset from cluster center, $\theta_{\perp}$ (\arcsec) \dotfill & 19.8  \\
Spin period, $P$ (ms)  \dotfill  & 3.003500835979(8)  \\
Orbital period, $P_b$ (days) \dotfill & 1.378005120(6)  \\
Time of ascending node passage, $T_{asc}$ (MJD) \dotfill & 58398.0011780(7) \\
Projected semi-major axis, $x$ (s) \dotfill & 1.251702(3)\\
Eccentricity, $e$ \dotfill & 5(4)$\times 10^{-6}$ \\
NS mass function, $f \ (M_{\odot})$ \dotfill  & 0.001108878(8)  \\
\hline
\enddata
\tablenotetext{a}{Numbers in parentheses are uncertainties in the last digits quoted.}
\end{deluxetable*} 

\section{Data-set and Data Reduction}
\label{sec:data}
We used deep and high-resolution \textrm{Hubble Space Telescope} (HST) images obtained with the {\rm Wide Field Camera 3} (WFC3) and the {\rm Advanced Camera for Surveys} (ACS) under GO 12605 (PI: Piotto), GO 10775 (PI: Sarajedini) and GO 10349 (PI: Lewin). The adopted data-set consists of a selection of images obtained with different filters covering a wavelengths range from near-UV to optical. The complete log of the observations is reported in Table~\ref{tab:obslog}.   

\begin{deluxetable*}{c|c|c|c}
\tablecaption{Observations Log\label{tab:obslog}}
\tablewidth{0pt}
\tablehead{
\colhead{Obs. ID} & \colhead{Camera} & \colhead{Filter} &  \colhead{Exposures}}
\startdata
         &  & F275W & 6 x 427 s \\
G0 12605 & WFC3/UVIS & F336W & 4 x 350 s      \\
         &  & F438W & 4 x 46 s     \\
\hline
GO 10775 & ACS/WFC & F606W & 1 x 7 s; 4 x 140 s     \\
         & & F814W & 1 x 7 s; 3 x 140 s     \\
\hline
         & & F435W & 1 x 120 s; 2 x 680 s     \\
GO 10349 & ACS/WFC & F625W & 1 x 20 s; 4 x 90 s      \\
         & & F658N & 1x 120 s; 1 x 690 s; 1 x 800 s    \\
\hline
\hline
\enddata
\end{deluxetable*}

The photometric analysis was performed using {\rm DAOPHOT IV} \citep{stetson87} and adopting the so-called ``UV-route'' described in \citealt{raso17} (see also \citealt{cadelano19}). First of all, we selected $\sim200$ bright stars to model the point spread function of each image. These models were then applied to all the sources detected at more than $5\sigma$ from the background level. Then, we created a master list of stars with objects detected in at least half the F275W images. At the corresponding positions of these stars, the photometric fit was forced in all the other frames by using {\rm DAOPHOT/ALLFRAME} \citep{stetson94}. By adopting such a near-UV master list, the crowding effects due to the presence of giants and turn-off stars are strongly mitigated and several blue stars like blue stragglers and white dwarfs are recovered. Finally, for each star we homogenized the magnitudes estimated in different images, and their weighted mean and standard deviation have been adopted as the star magnitude and its related photometric error. The instrumental magnitudes were calibrated to the VEGAMAG photometric system by cross-correlation with the publicly available catalogs of ``{\it The Hubble Space Telescope UV Legacy Survey of Galactic globular clusters}'' \citep{piotto15}, of the ``{\it ACS Globular Cluster Survey}'' \citep{sarajedini07} or by using appropriate filter zero points and aperture corrections as described in \citet{bohlin16}.

The stellar detector positions were corrected for geometric distortion effects following the prescriptions by \citet{meurer03} and \citet{bellini11}. Finally, the corrected positions were converted to the absolute coordinate system ($\alpha$,$\delta$) using the stars in common with the Gaia DR2 catalog \citep{gaia18}. The coordinate system of this catalog is based on the International Celestial Reference System, which allows an accurate comparison with the MSP positions derived from timing using solar system ephemerids, since the latter are referenced to the same celestial system. The resulting combined $1\sigma$ astrometric uncertainty is $\lesssim 0.1 \arcsec$.

\section{The companion to M13F}
\label{sec:comp}
\subsection{Identification of the companion star}

\begin{figure}
\centering
\includegraphics[scale=0.4]{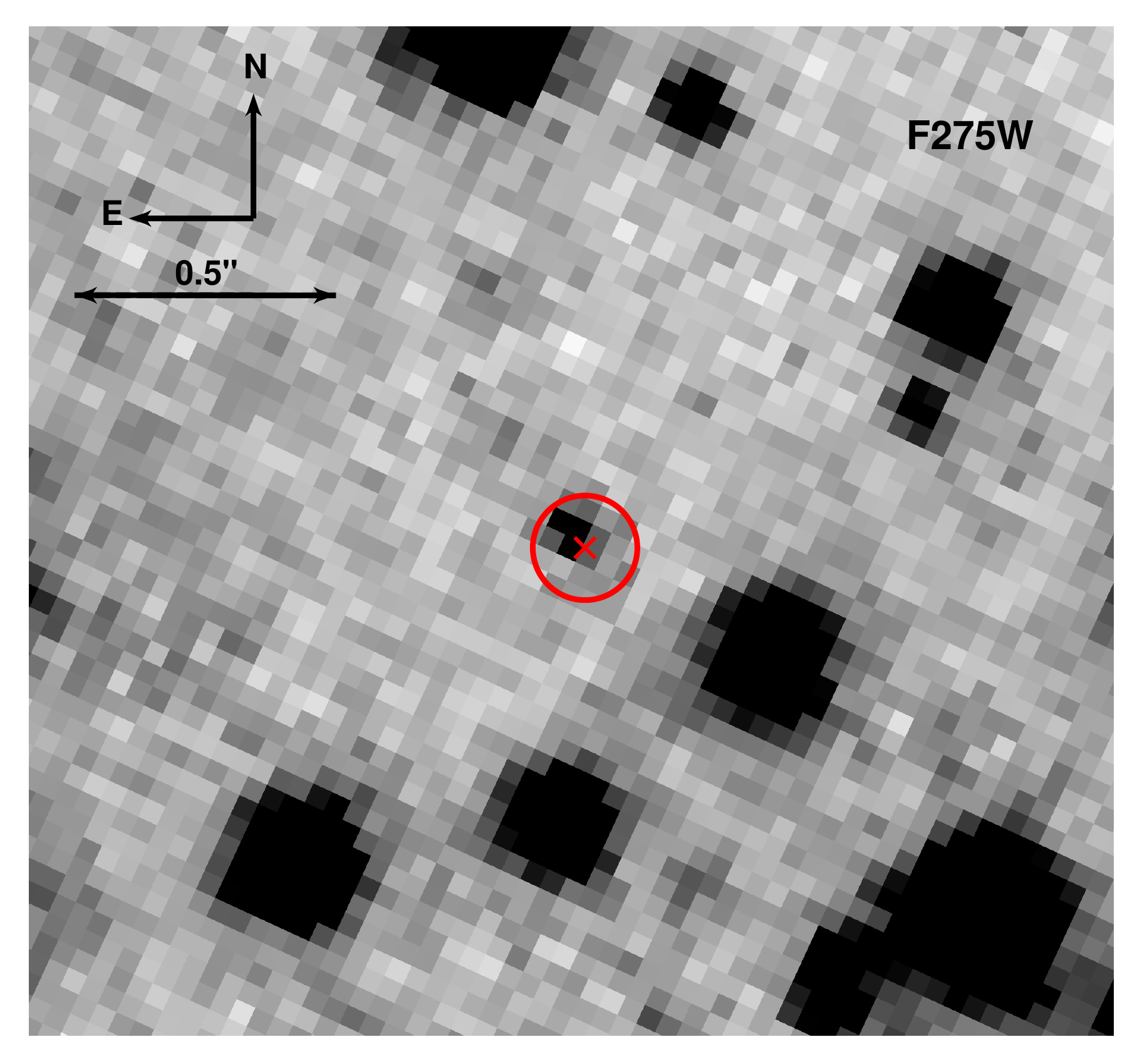}
\includegraphics[scale=0.4]{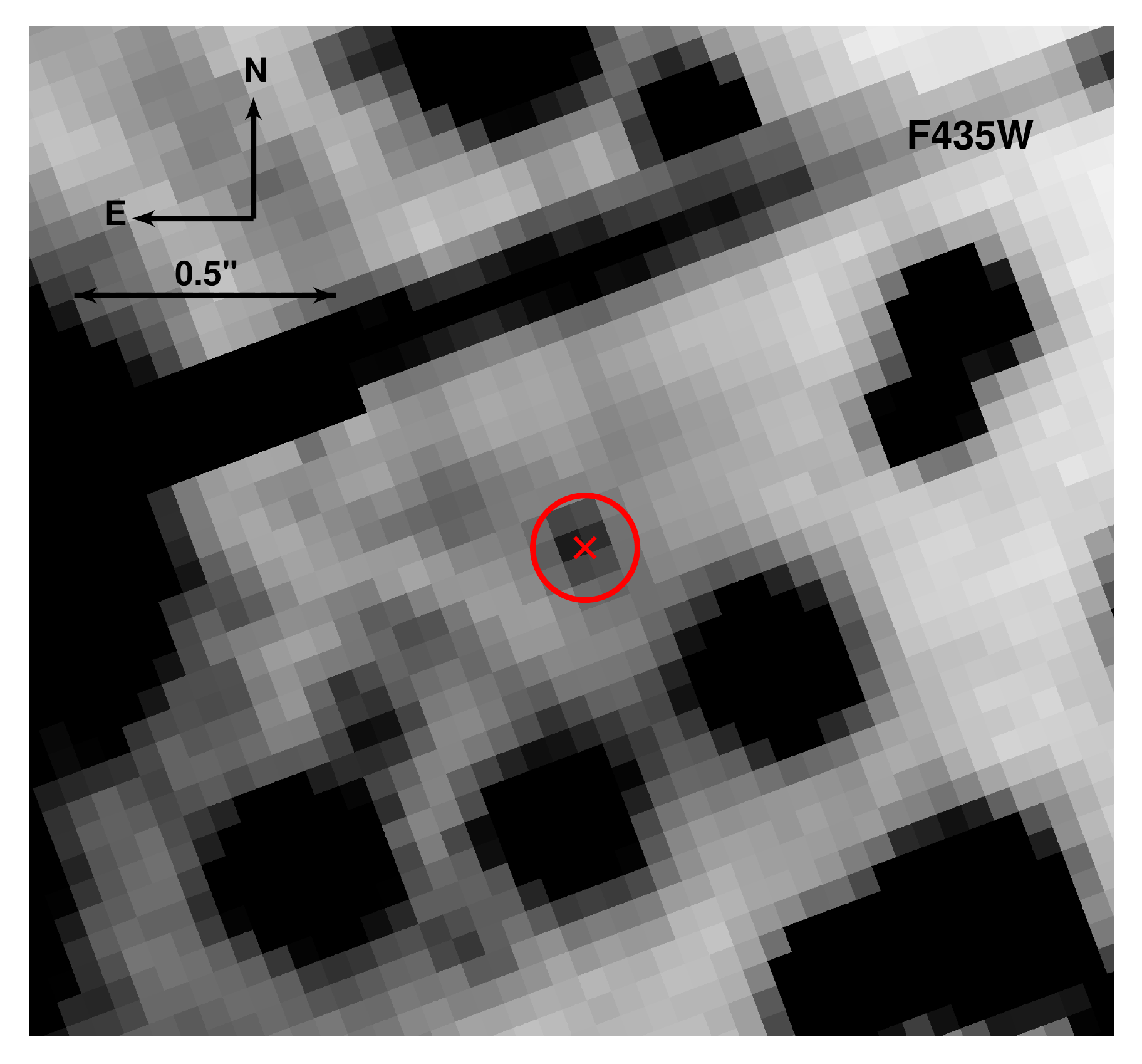}

\caption{{\it Left Panel:} $2\arcsec \times 2\arcsec$ region surrounding the position of M13F in a F275W image. The red cross is centered on the MSP position while the red circle has a radius of $0.1\arcsec$.  The only star located within this circle is the identified companion to M13F. {\it Right Panel:} same as in the left panel, but for a F435W image.}
\label{fig:chart}
\end{figure}

\begin{figure}
\centering
\includegraphics[scale=0.6]{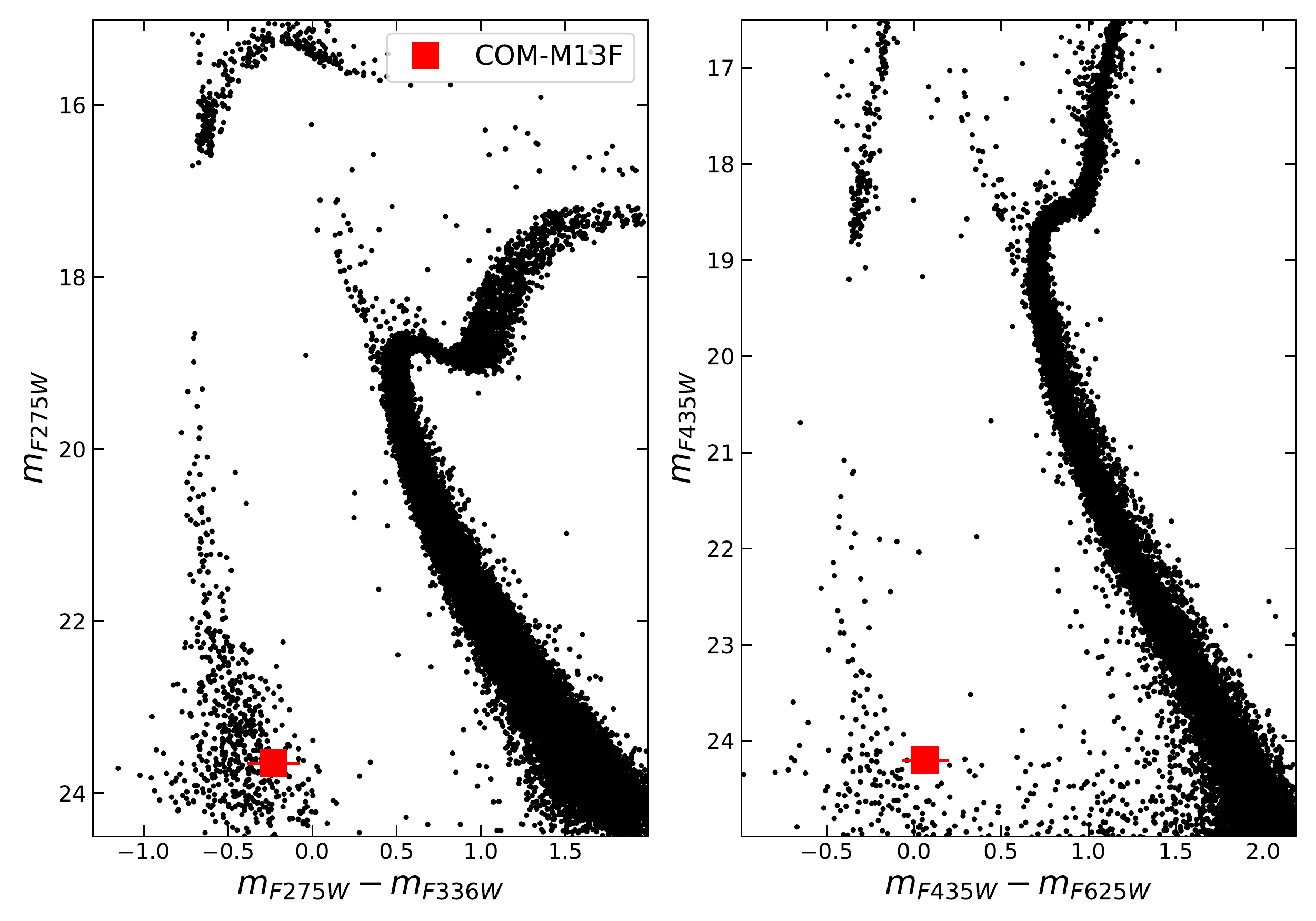}
\caption{{\it Left panel:} ($m_{F275W}$, $m_{F275W}-m_{F336W}$) near-UV CMD of M13. {\it Right panel:} ($m_{F435W}$, $m_{F435W}-m_{F625W}$) optical CMD of M13. In both panels, the position of the companion star to M13F is highlighted with a red square {and the error bars correspond to the $1\sigma$ confidence level uncertainties.}}
\label{fig:cmd}
\end{figure}

We searched for the optical counterparts to all the binary MSPs in the cluster by carefully analyzing all the stars located within a $1\arcsec \times 1 \arcsec$ region centered on the radio positions. For each of the candidate stars, we analyzed its position in the color-magnitude diagrams (CMDs) and investigated the presence of photometric variability, possibly associated to the binary orbital period. Unfortunately, no interesting candidates have been discovered at the positions of M13B, M13D and M13E. The detection of the counterparts to these systems was likely hampered by the presence of saturated or very bright stars close to their positions. On the other hand, we identified a promising candidate for the binary system M13F. In fact, at a distance of only $0.02 \arcsec$ from the radio position, we discovered an extremely faint and blue star whose finding chart is reported in Figure~\ref{fig:chart}. Despite its very low luminosity, this object is detectable in all the available filters, with the only exception of the F658N one. Its magnitudes {and $1\sigma$ uncertainties} are: $m_{F275W}=23.65\pm0.10$, $m_{F336W}=23.9\pm0.1$, $m_{F438W}=24.2\pm0.2$, $m_{F435W}=24.20\pm0.02$, $m_{F606W}=24.35\pm0.07$, $m_{F625W}=24.1\pm0.1$ and $m_{F814W}=24.2\pm0.1$, while for the F658N filter we derived a lower limit of $m_{F658N}>24.2$. As shown in Figure~\ref{fig:cmd}, these magnitudes place the candidate companion star along the red side of the WD cooling sequence, in a region compatible with the expected location of WDs with a He core. The excellent agreement between the radio and optical positions and its peculiar location in the CMDs allow us to safely conclude that the detected object is the companion star to M13F.

\subsection{Comparison with binary evolution models}
\label{sec:comparison}
\begin{figure}
\centering
\includegraphics[scale=0.6]{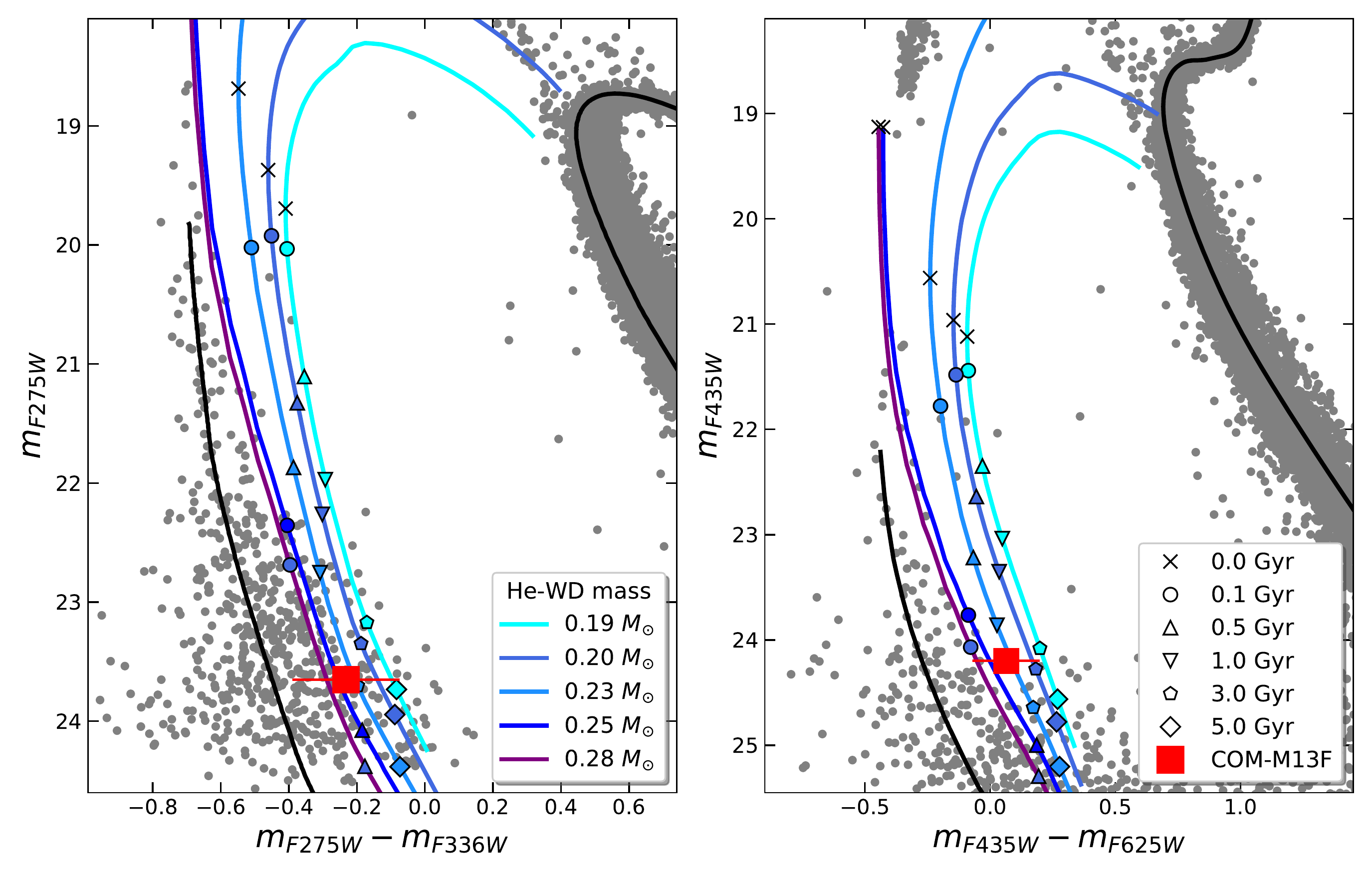}
    \caption{Same as in Figure~\ref{fig:cmd} but zoomed on the WD region. The black curves are an isochrone reproducing a 13 Gyr old stellar population and a cooling track for a CO-WD with a mass of $0.55 \ M_{\odot}$. The other curves are He-WD tracks with masses, from left to right, of $0.28 \ M_{\odot}$, $0.25 \ M_{\odot}$, $0.23 \ M_{\odot}$, $0.20 \ M_{\odot}$ and $0.19 \ M_{\odot}$ computed similarly as in \citet{istrate14a,istrate16}.  Points at different cooling ages are highlighted with different symbols as reported in the right panel legend.}
\label{fig:cmdwd}
\end{figure}

In the previous section we confirmed that M13F is orbiting a WD. In order to get insights on the properties of the companion star, it is useful to compare its observed magnitudes with binary evolution models. First, in order to properly compare the observed and theoretical frames, we checked the accuracy of the photometric calibration by comparing the observed cluster sequences in the CMDs, such as the main sequence and the CO-WD sequence, with isochrones and CO-WD cooling tracks. We generated from the {\it BaSTI} database \citep{pietrinferni04,pietrinferni06,salaris10} an isochrone reproducing a 13 Gyr old stellar population \citep{dotter10} with a metallicity $[Fe/H]=-1.62$ and $[\alpha/Fe]=0.2$, together with a cooling track for a CO-WD with a canonical mass $M=0.55 \ M_{\odot}$. Absolute magnitudes were converted to be observed frame by adopting a distance modulus $(m-M)_0=14.43$, a color excess $E(B-V)=0.02$ \citep{ferraro99,dotter10} and appropriate extinction coefficients calculated following the prescriptions by \citet{cardelli89,odonnell94}. The two curves are shown in Figure~\ref{fig:cmdwd}. The excellent agreement between the models and the observed sequences confirms the accuracy of the photometric calibration and of the adopted cluster parameters. {Moreover, the position of the companion to M13F with respect to the CO-WD cooling track also suggests that it is likely a WD with a He core, as expected from the canonical formation scenario of MSPs.}

{To derive the companion properties, we computed binary evolutionary models using  the open source stellar evolution code MESA{\citep{paxton2011, paxton2013, paxton2015, paxton2018}, version 12115, in a similar fashion to \citet{istrate14a,istrate16}}. The initial binary parameters consist of a 1.4 $M_{\odot}$ NS, treated as point mass, and a 1.1 $M_{\odot}$ donor star with a metallicity of $Z=0.0005$, compatible with that of the cluster. The new evolutionary models will be soon publicly available (Istrate et al. 2020, in prep).}
  The resulting tracks span a He-WD mass range of $0.18 \ M_{\odot}-0.4 \ M_{\odot}$, a surface temperature range of $6000 \ K - 21000 \ K$ and cooling ages\footnote{The WD cooling age is defined, according to \citet{istrate16}, as the time passed since the proto-WD reached the maximum surface temperature along the evolutionary track.} up to the cluster age. {Theoretical bolometric luminosities were transformed to HST magnitudes by using the Astrolib PySynphot package \citep{pysynphot}\footnote{\url{ https://pysynphot.readthedocs.io/en/latest/}} and WD spectra templates by \citet[][see also \citealp{tassoul90,tremblay09}]{koester10}}. A selection of these evolutionary tracks is shown in Figure~\ref{fig:cmdwd}, where we can qualitatively infer that the track better reproducing the companion CMD positions is the one corresponding to a mass of $0.23 \ M_{\odot}$.  

\begin{figure}
\centering
\includegraphics[scale=0.4]{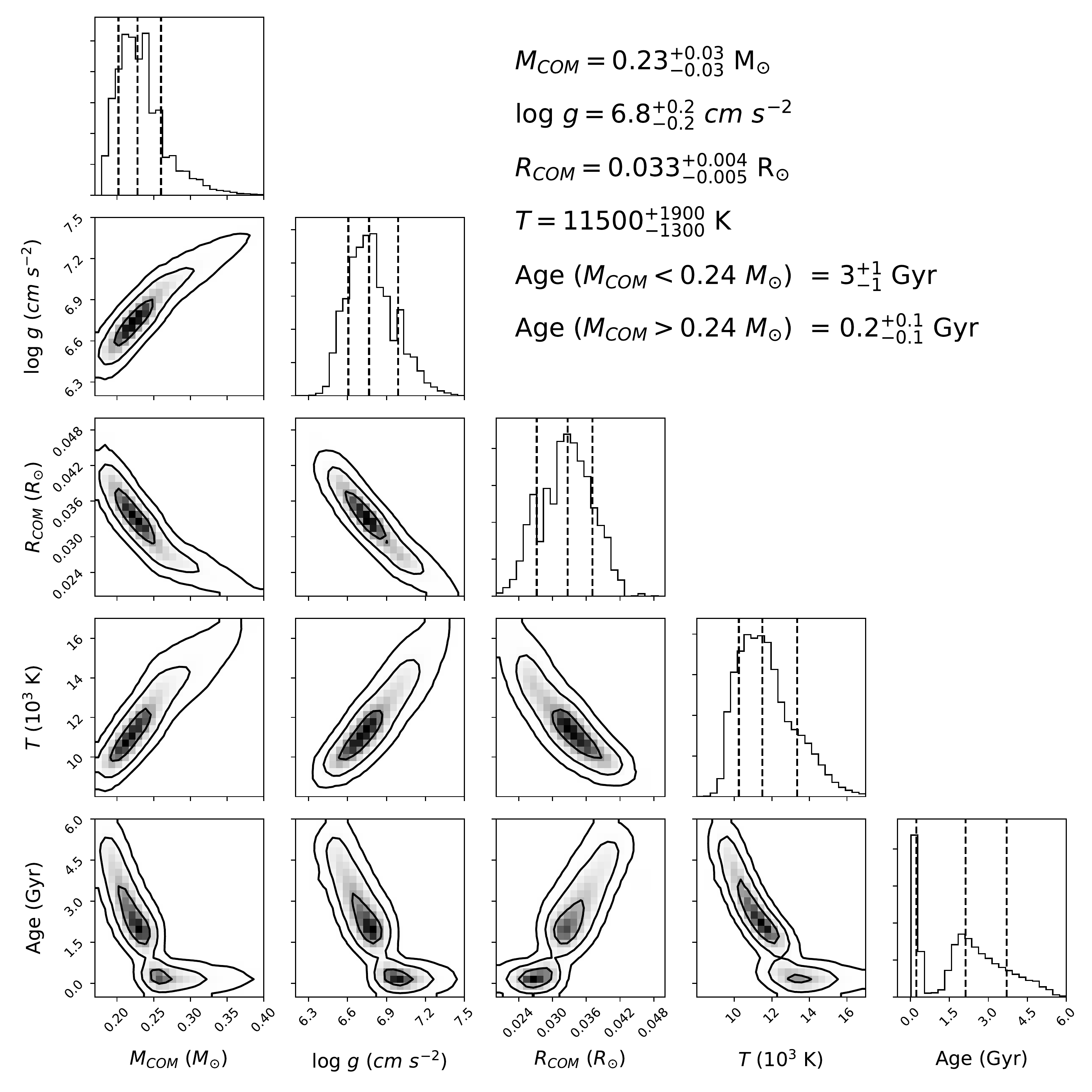}
\caption{Constraints on the mass, surface gravity, radius, surface temperature and cooling age of the companion star to M13F. The 1D histograms show the likelihood weighted distributions for each of the  parameters. The three vertical dashed lines for each 1D histogram correspond, from left to right, to the 0.16, 0.5, and 0.84 quantiles, respectively. The contours in the 2D histograms correspond to $1\sigma$, $2\sigma$ and $3\sigma$ levels. The text at the top reports the derived values for each parameter.}
\label{fig:corner}
\end{figure}

In order to get a quantitative derivation of the companion physical properties (such as its mass, radius, age, surface gravity and temperature) we implemented an approach similar to that described in \citet{cadelano19}. We defined a logarithmic likelihood $\mathscr{L}$ to quantify the probability of each point of each evolutionary track to reproduce the  observed companion magnitudes, as follows:
\begin{equation}
\label{like1}
\ln \mathscr{L} = -\frac{1}{2} \sum_{f} \frac{(m_{f} - \tilde{m}_{f})^2}{\delta_f^2} + \ln (2 \pi \delta_f^2) 
\end{equation}
where the index $f$ runs through all the available filters, $m_{f}$ and $\tilde{m}_{f}$ are the observed and the model magnitudes in a given filter $f$, respectively, and $\delta_f$ is the uncertainty on the observed magnitude. The latter term also takes into account a 0.1 mag uncertainty on the cluster distance modulus and a 0.01 mag uncertainty on the cluster color excess. {The likelihood $\mathscr{L}$ was also forced to zero wherever the predicted F658N magnitude was smaller than the lower limit derived in the previous section}.
The resulting likelihood-weighted 1D and 2D histograms are presented in the corner plot\footnote{\url{https://corner.readthedocs.io/en/latest/} \citep{foreman16}} of  Figure~\ref{fig:corner}. For each of the WD parameters, we derived the best values and related uncertainties as the 0.16, 0.5 and 0.84 quantiles of the distributions. All these results are also listed in Table~\ref{tab:comp}.

\begin{deluxetable*}{cc}
\tablecaption{Derived properties of the companion to M13F \label{tab:comp}}
\tablewidth{0pt}
\tablehead{
\colhead{Param.} & \colhead{Value} }
\startdata
$M_{COM} \ (M_{\odot})$ \dotfill & $0.23\pm0.03$    \\
$\log g$ (cm/s$^{2}$) \dotfill & $6.8\pm 0.2$ \\
$R_{COM} \ (R_{\odot})$ \dotfill & $0.033^{+0.004}_{-0.005}$    \\
$T_{eff}$ (K) \dotfill & $11500^{+1900}_{-1300}$ \\
Cooling Age (Gyr) \dotfill & $3\pm1$ or $0.2\pm0.1$ \\
Proto-WD Age (Myr) \dotfill & $200^{+400}_{-100}$ \\
\hline
\hline
\enddata
\tablecomments{From top to bottom: companion mass, surface gravity, radius, surface temperature, cooling age and proto-WD age.}
\end{deluxetable*}

Results show that the companion to M13F is indeed a low-mass He-WD with a mass around $0.23 \ M_{\odot}$. As it can be seen in Figure~\ref{fig:corner}, all the companion parameters have been firmly constrained with the exception of the cooling age. Indeed, the 1D histogram of the cooling ages show a clear bi-modal distribution. This feature is the result of a dichotomy in the cooling timescales due to the occurrence of {diffusion induced hydrogen shell flashes }{in the envelope of proto-WD with mass $M\gtrsim M_{\mathrm{flash}}\sim0.2\; M_{\odot}$}\footnote{Please note that this critical value strongly depends on the metallicity and physics of diffusion \citep{istrate16}}. {The systems which experience such flashes will enter the cooling tracks with a thin hydrogen envelope. The WDs with $M<M_{\mathrm{flash}}$ experience stable hydrogen shell burning  during the proto-WD  and enter the cooling track with a thick hydrogen envelope. As the cooling timescale depends primarily on the mass of the hydrogen envelope, a cooling  dichotomy will be observed \citep[see e.g.][]{althaus01a,althaus01b,istrate14b,istrate16}}. This dichotomy clearly applies to our results. Indeed, our models predict that the minimum mass for the flashes to occur at the cluster metallicity is around $0.24 \ M_{\odot}$. A closer inspection to the 1D and 2D likelihood distributions of the cooling ages (Figure~\ref{fig:corner}) reveals a prominent narrow peak centered around 0.3 Gyr and a long tail extending up to $\sim5$ Gyr. The narrow peak is associated with masses larger than $0.24 \ M_{\odot}$, which experience hydrogen flashes (fast cooling), while the tail is associated with masses lower than $0.24 \  M_{\odot}$, which burns hydrogen stably (slow cooling). All this hampers a proper determination of the system age and we therefore report both the cooling ages in the figure and in Table~\ref{tab:comp}. {Selecting only the narrow peak of the age likelihood distribution, the corresponding companion mass is $0.26^{+0.04}_{-0.01} \ M_{\odot}$. On the other hand, the extended tail of the age likelihood distribution implies a mass of $0.22\pm0.02 \ M_{\odot}$.} Unfortunately, not even the pulsar spin-down age can be used to better constrain the system age, since its value has been proven to be highly unreliable \citep[e.g.][]{tauris12a,tauris12b} and also depends on the intrinsic pulsar spin-down rate, which cannot be easily determined for MSPs in GCs due to the contamination by the acceleration induced by the cluster potential \citep[e.g.][]{prager17}.

Although the data-set is composed of several multi-epoch images, we found no evidence of photometric variability. While this could be due to the very poor and random orbital period coverage provided by the available data-set, we stress that He WD companions only rarely show variability, which is usually due to pulsations (global stellar oscillations) of the WD itself \citep[e.g.][]{maxted13,kilic15,antoniadis16wd,parson20}. Indeed, heating of the stellar side exposed to the MSP and/or tidal distortions due to the NS tidal field are negligible , at odds with the case of non-degenerate and tidally-locked companions stars \citep[e.g.][]{pallanca10,pallanca14,cadelano15_m71}, {although exceptions exist \citep[e.g.][]{edmonds01,kaplan12}}.

\begin{figure}
\centering
\includegraphics[scale=0.75]{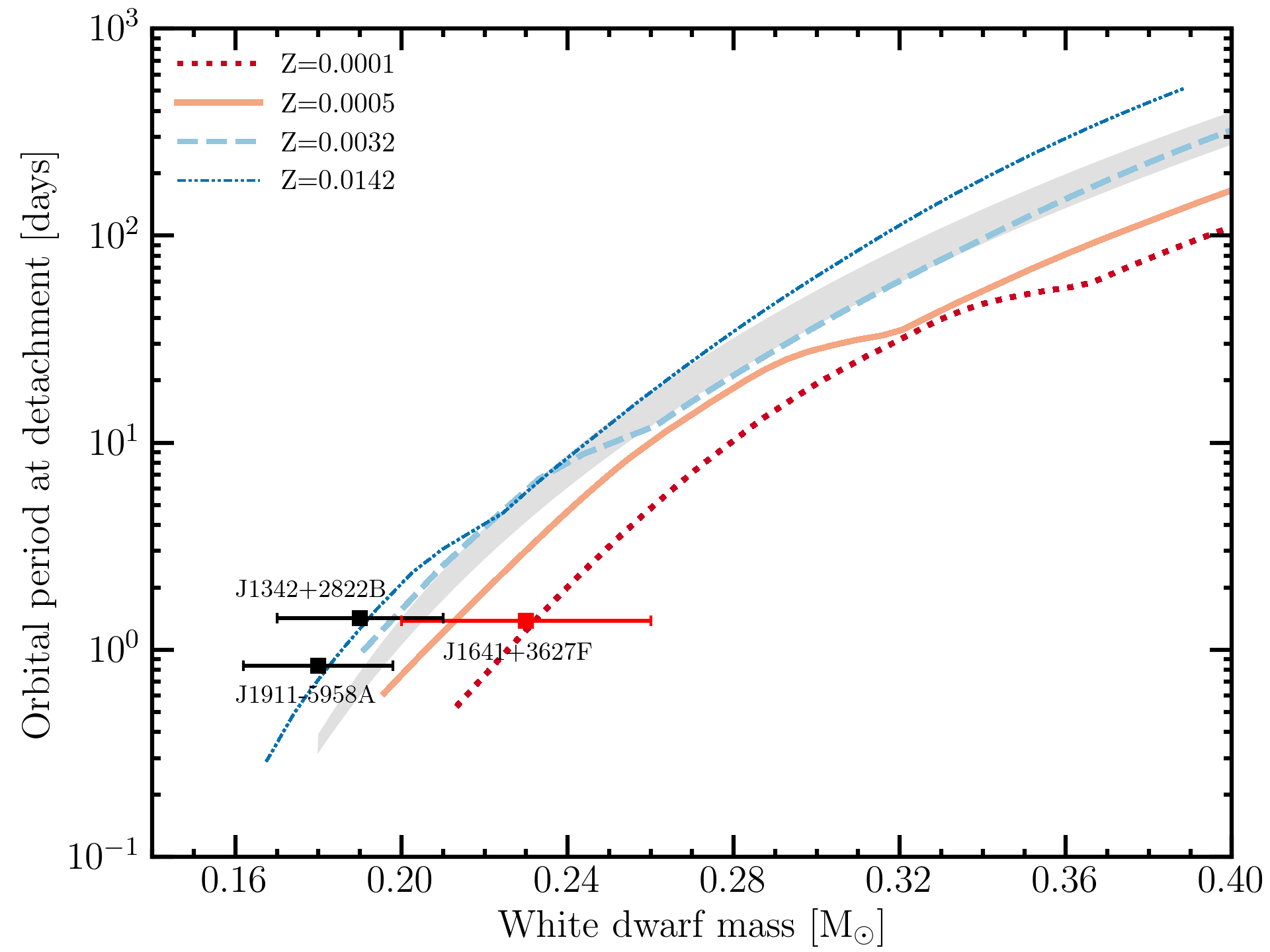}
\caption{{Orbital period at the end of the mass-transfer phase versus the mass of the proto-WD. The various lines represent theoretical mass-period relations for Z=0.0001, Z=0.0005, Z=0.0032 and Z=0.0142 as predicted by binary evolution models (Istrate et. al in prep.)}. The grey region represents the  fitted mass-period relation by \citet{tauris99}. The orange solid line at Z=0.0005 is representative for the metallicity of M13. The red square reports the position of the companion to M13F. We also included the positions of the companions to J1342$+$2822B in M3  \citep{cadelano19} and J1911$-$5958A in NGC6752  \citep{corongiu12}, since the two clusters share approximately the same metallicity of M13.}
\label{fig:correlation}
\end{figure}

The companion mass here derived is based on the comparison between the WD observed and predicted optical magnitudes and thus does not take into account the orbital properties of the system derived through radio timing. However, for He-WDs formed through the stable mass-transfer channel, there is a very well known and tight correlation between the binary orbital period and the mass of the proto-WD at the epoch of binary detachment \citep[e.g.][]{sav87,joss1987,rappaport1995,tauris99,lin2011,istrate14b}. Such a correlation, which mainly depends on the companion star metallicity and, to a lesser extent, on other stellar parameters (e.g. mixing length parameter, initial companion mass), has been also confirmed through observations \citep[e.g.][]{corongiu12}. {The theoretical mass-period relation for various metallicities is shown in  Figure~\ref{fig:correlation}. One should note that compared to field stars, GC exhibit helium and $\alpha$ elements enhancement which  were ignored in this work, but might influence the mass-period relation at a given metallicity. These effects will be studied in a future work.} At the  orbital period of M13F and cluster metallicity ($Z\approx0.0005$), the predicted mass for the forming He-WD is $\sim 0.21 \ M_{\odot}$, in agreement with our results. Here it is assumed that the WD did not lose a significant amount of mass during the proto-WD stage and that the current orbital period is almost unchanged with respect to that at the epoch of the binary detachment.

\subsection{Constraints on the NS mass}

\begin{figure}
\centering
\includegraphics[scale=0.45]{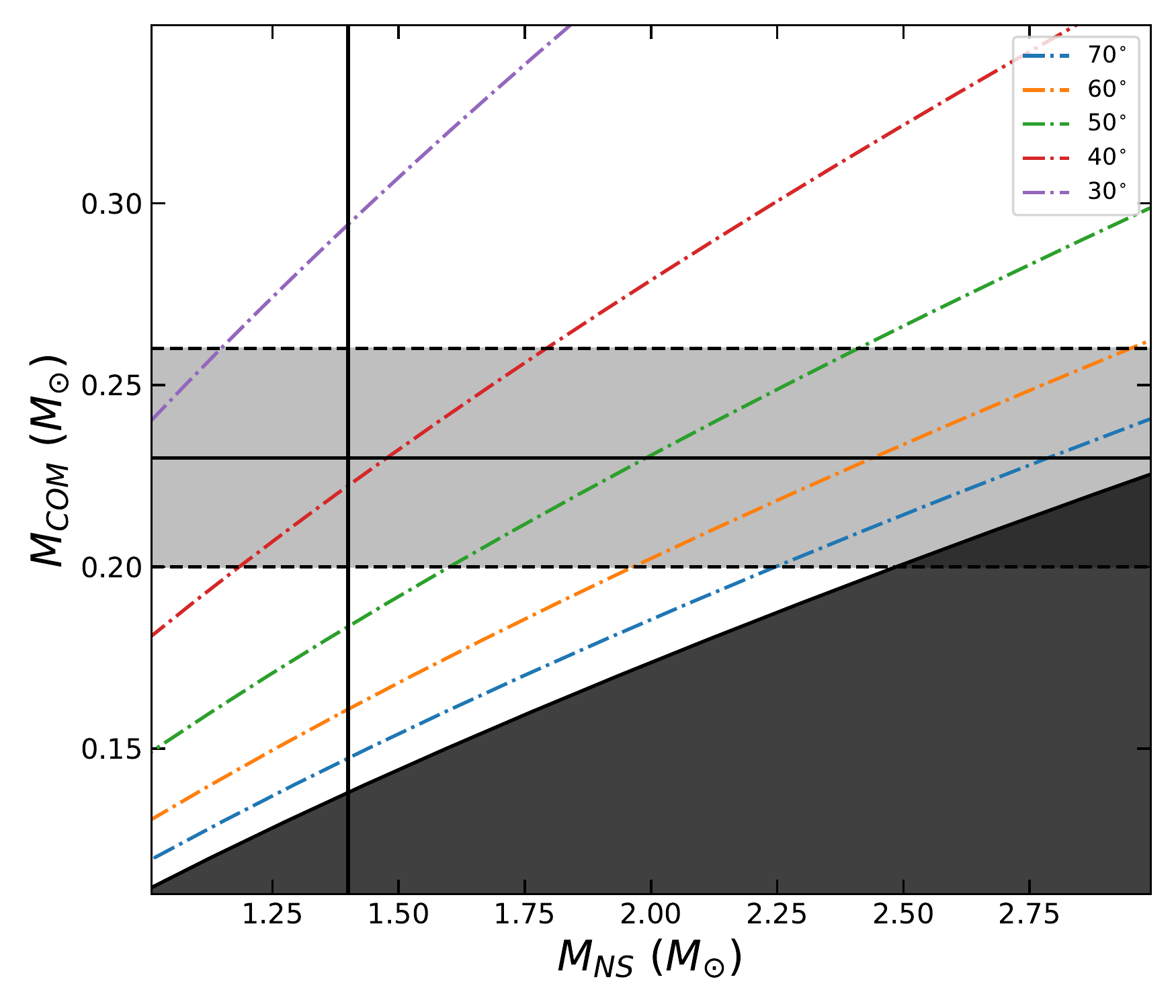}
\includegraphics[scale=0.4]{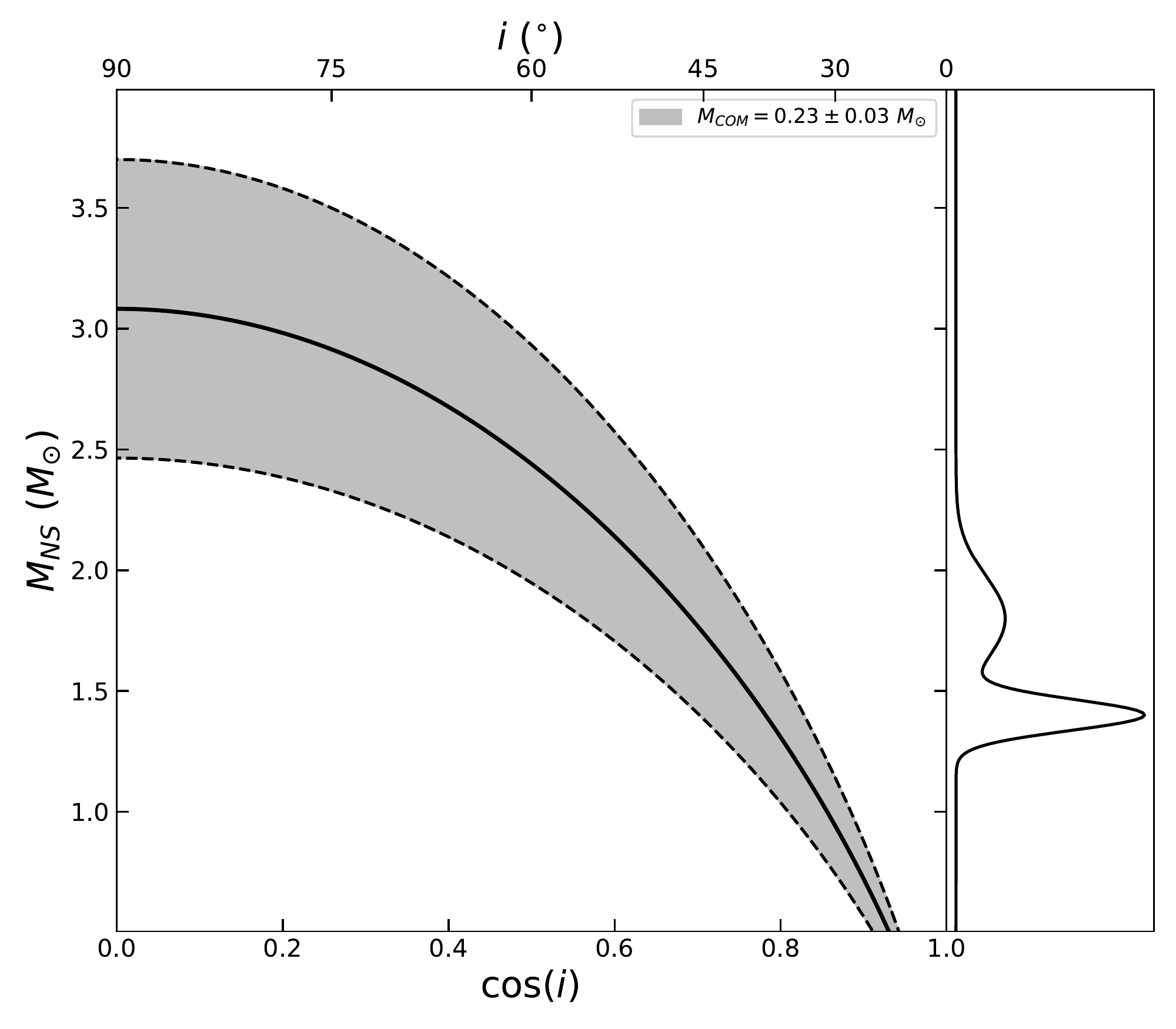}
\caption{{\it Left panel:} companion mass as a function of the NS mass. The solid horizontal line marks the best-fit value of the companion mass ($M=0.23 \ M_{\odot}$) and the light-grey region delimited by the two horizontal dashed lines marks its estimated uncertainty. The dark gray shaded area is the region forbidden by the binary mass function (see equation~\ref{mfunc}), while the dot-dashed colored lines are the curve obtained assuming different inclination angles. The vertical line represents a canonical NS mass of $1.4 \ M_{\odot}$. {\it Right panel:} NS mass as a function of the cosine of the orbital inclination angle, for the estimated mass of the companion star: the solid curve marks the combination of values allowed by the best-fit value of the companion mass ($0.23 \ M_{\odot}$), while the light-grey region delimited by the two  dashed curves marks the combinations allowed within the uncertainty (see the legend). {The right edge panel shows, as reference, the NS mass distribution empirically derived by \citet{antoniadis16}.}}
\label{fig:mpsrinc}
\end{figure}

\begin{figure}
\centering
\includegraphics[scale=0.5]{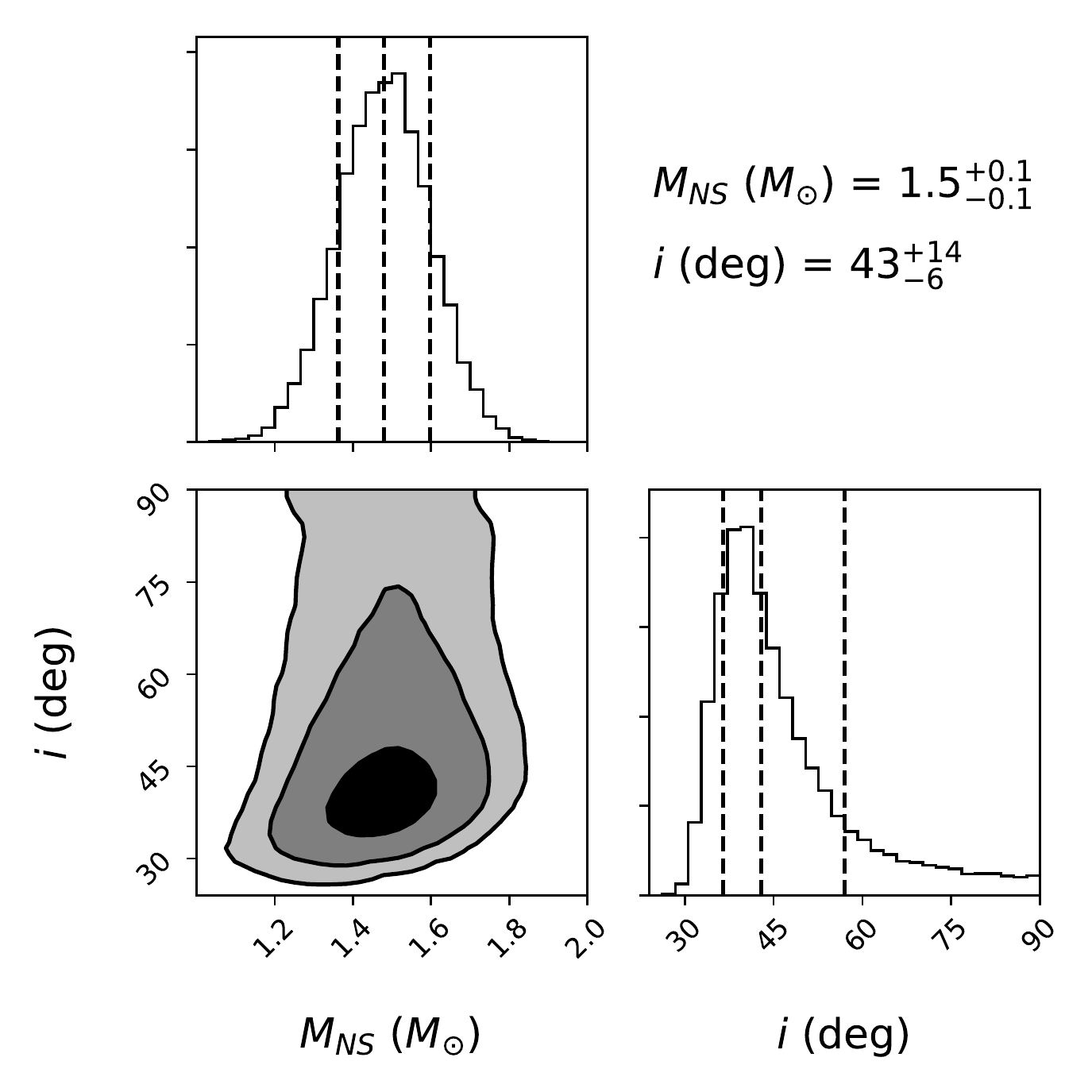}
\caption{{Corner plots showing the 1D and 2D projections of the posterior probability distribution of the NS mass and inclination angle of  M13F. The 1D-histograms are the marginalized probability distributions and the dashed lines corresponds to their $16^{th}$,$50^{th}$,$84^{th}$ percentiles. The bottom left panel is the joint 2D posterior probability distribution and the contours corresponds to $1\sigma$, $2\sigma$ and $3\sigma$ confidence levels.}}
\label{fig:mcmc_mpsr}
\end{figure}

The determination of the companion mass together with the binary orbital parameters derived through the radio timing analysis allows us to place constraints on the NS mass. In fact, the masses of the binary components can be expressed as a function of the orbital parameters through the NS mass function:
\begin{equation}
\label{mfunc}
\frac{(M_{COM} \sin i)^3}{(M_{NS}+M_{COM})^2}= \frac{4\pi^2 x^3}{GP_{ORB}^2}
\end{equation}

where $M_{COM}$ and $M_{NS}$ are the companion and the NS mass, respectively, $i$ is the orbital inclination angle, $G$ the gravitational constant, $x$ the projected semi-major axis and $P_{ORB}$ the orbital period. The right-hand side of this equation depends exclusively on the binary orbital parameters (see Table~\ref{tab:m13f}) and thus its value is very well constrained: $f=0.001108878\pm 0.000000008 \ M_{\odot}$. On the other hand, the left-hand side of the equation contains the measured companion mass and two completely unknown quantities: the NS mass and the orbital inclination angle. {Figure~\ref{fig:mpsrinc} shows the NS mass predicted by equation~\ref{mfunc} as a function of both the companion mass and the orbital inclination angle. We find that a $0.23\pm0.03 \ M_{\odot}$ companion star could in principle imply the presence of a very massive NS.} In fact, Figure~\ref{fig:mpsrinc} reveals that if M13F is observed almost edge-on ($i\gtrsim70^{\circ}$), than the NS mass should be $\gtrsim2.4 \ M_{\odot}$. More precisely, the maximum NS mass, corresponding to $i=90^{\circ}$ is $M_{NS,max}=3.1\pm0.6 \ M_{\odot}$. Such value is larger than ever measured for any massive NS and larger than the maximum sustainable mass predicted by most of the theoretical equations of state \citep[e.g.][]{ozel16}. Therefore, it is unlikely that this system is observed at very large inclination angle. 
On the other hand, {assuming that binaries are randomly inclined with respect to the observer, i.e. assuming a flat distribution of $\cos i$, we find a median NS mass (corresponding to $i=60^{\circ}$) of $M_{NS,med}=2.4\pm0.5 \ M_{\odot}$ (see right panel of Figure~\ref{fig:mpsrinc})}. Such value would place M13F on the highest-mass side of the known NS mass distribution \citep[e.g.][]{antoniadis16,cromartie20}. {All this suggests that either M13F is observed at quite small inclination angles ($i\lesssim40^{\circ}$) or it hosts a massive NS.
Indeed, under the assumption of 
a flat distribution of $\cos i$ , there is a $\sim70\%$ of probability that the compact object mass is larger than $1.6 \ M_{\odot}$, thus favoring the case of a massive NS. 
Finally, we further investigate the possibility of having a standard NS mass coupled with a low inclination angle. To this aim, we used a Monte Carlo Markov Chain sampler \citep{emcee_v3} to explore the combination of NS masses and orbital inclination angles able to reproduce the observed mass-function. We defined a standard Gaussian likelihood function to minimize the difference between the left and right side of equation~\ref{mfunc}. We assumed an uniform prior on the distribution of $\cos{i}$ and also a prior on the NS mass distribution following that empirically derived by \citet{antoniadis16}, which is roughly a double Gaussian with a main component centered at $1.4 \ M_{\odot}$ and with a dispersion of about  $0.1 \ M_{\odot}$, and a secondary component centered at $1.8 \ M_{\odot}$ and with a dispersion of about $0.2 \ M_{\odot}$. The posterior distribution is shown in Figure~\ref{fig:mcmc_mpsr} and the results based on the $16^{th}$,$50^{th}$,$84^{th}$ percentiles show that M13F could host a NS with an almost standard mass of $1.5 \pm 0.1 \ M_{\odot}$ if orbiting with an inclination angle $i=43^{+15}_{-6}$ degrees.}



\section{Conclusions} \label{sec:conc}

PSR J1641+3627F is a binary MSP recently discovered in the GC M13 by observations with the FAST radio-telescope \citep{wang20}. Its timing analysis revealed a 1.4 day circular orbit and a NS mass function implying a minimum and median companion mass of only $0.13 \ M_{\odot}$ and $0.16 \ M_{\odot}$, respectively, under the assumption of a NS having with a standard mass of $1.4 \ M_{\odot}$. However, state-of-the-art binary evolution models suggest that such low-mass companions are unlikely to be produced in a intermediate-low metallicity cluster such as M13. {This is mainly due to the fact that  lower metallicity stars are more compact for the same He-core mass compared to higher metallicities \citep{istrate16}.} This suggests that M13F could host a non-canonical companion or, alternatively,  that the system is observed at nearly face-on orbits or that it hosts a high-mass NS. To shed light on this, we used a combination of near-UV and optical observations obtained with HST to identify the companion star. At a distance of only $0.02\arcsec$ from the radio MSP position, we identified a faint and blue object located on the red side of the CMD region occupied by the cluster WD cooling sequences and thus compatible with the expected position of He-WDs. {Our study therefore allowed us to exclude the first possibility and conclude that the companion to M13F is a canonical He-WD.} We exploited the HST multi-band photometry also to constrain the companion properties from the comparison with {binary evolutionary models computed for M13 metallicity following the prescriptions in \citet{istrate16}.}  We found that the companion to M13F likely has a mass  $M_{COM}=0.23\pm0.03 \ M_{\odot}$. It is important to stress that this value is model-dependent, but different assumptions about the binary and stellar evolution physics are expected to lead to only slightly different results.


{Although our analysis does not allow to break the degeneracy between the inclination angle and the NS mass, we could at least reduce the range of possibilities by combining the derived companion mass with the binary orbital parameters. We find a maximum NS mass}
 (corresponding to an edge-on orbit, $i=90^{\circ}$) of $M_{NS,max}=3.1\pm0.6 \ M_{\odot}$ and a median NS mass (corresponding to $i=60^{\circ}$) of $M_{NS,med}=2.4\pm0.05 \ M_{\odot}$. These high values suggest that M13F is unlikely to be observed edge-on. Therefore we conclude that either M13F hosts a canonical NS with a mass $\sim1.4 \ M_{\odot}$ and is observed at nearly face-on inclination angles around $40^{\circ}$ or its NS is part of the growing class of high-mass NSs. 

To break the degeneracy between a face-on orbit and a high-mass NS, independent measurements of the companion mass and possibly of the orbital inclination are needed. 
{These quantities could in principle be constrained, for example, through the radio detection of a Shapiro delay \citep[e.g.][]{corongiu12,cromartie20}, in particular its orthometric amplitude parameter $h_3$ \citep[see][]{freire10}. However, given the extreme faintness of M13F, even in FAST data, it is unlikely that such a detection can be made in the near future.} Alternatively, the mass ratio between the two binary component could be measured through spectroscopic observations that however, due to the companion low-luminosity, are beyond the capabilities of the current generation of optical telescopes. 

Unfortunately, the wavelength coverage and sampling of our observations is not enough to perform a reliable SED fitting of the observed magnitudes with WD spectral templates. Although this technique could be useful to obtain an independent companion mass measurement, additional observations at UV wavelengths ($\lambda<2700 \AA$) are necessary to properly constrain the surface gravity and radius of a WD by using this method. However, multi-band and high quality observations sampling the whole wavelength range from UV to optical are indeed feasible  and could confirm in the future the companion mass here derived.

\acknowledgments
 {We thank the anonymous referee for useful comments that improved the presentation of the results}. This paper is part of the project Cosmic-Lab (``Globular Clusters as Cosmic Laboratories'') at the Physics and Astronomy Department of the Bologna University (see the web page: \url{http://www.cosmic-lab.eu/Cosmic-Lab/Home.html}). The research is funded by the project Light-on-Dark granted by MIUR through PRIN2017 contract (PI:Ferraro).\\
Based on observations with the NASA/ESA Hubble Space Telescope , obtained at the Space Telescope Science Institute, which is operated by AURA, Inc., under NASA contract NAS 5-26555. \\
This work has made use of  data from the European Space Agency (ESA) mission Gaia (\url{https://www.cosmos.esa.int/gaia}), processed by the Gaia Data Processing and Analysis Consortium (DPAC, \url{https://www.cosmos.esa.int/web/gaia/dpac/consortium}). A.G.I  acknowledges support from the Netherlands Organisation for Scientific Research (NWO).

\vspace{5mm} \facilities{HST(WFC3, ACS)} \software{DAOPHOT
  \citep{stetson87}, DAOPHOT/ALLFRAME \citep{stetson94}, {\rm emcee} \citep{foreman19}, {\rm corner.py} \citep{foreman16},  MESA \citep[v12115;][]{paxton2011,paxton2013,paxton2015,paxton2018}, PySynphot \citep{pysynphot} }

\bibliography{main}{}

\begin{thebibliography}{}
\expandafter\ifx\csname natexlab\endcsname\relax\def\natexlab#1{#1}\fi
\providecommand{\url}[1]{\href{#1}{#1}}
\providecommand{\dodoi}[1]{doi:~\href{http://doi.org/#1}{\nolinkurl{#1}}}
\providecommand{\doeprint}[1]{\href{http://ascl.net/#1}{\nolinkurl{http://ascl.net/#1}}}
\providecommand{\doarXiv}[1]{\href{https://arxiv.org/abs/#1}{\nolinkurl{https://arxiv.org/abs/#1}}}

\bibitem[{{Abbate} {et~al.}(2019){Abbate}, {Possenti}, {Colpi}, \&
  {Spera}}]{abbate19}
{Abbate}, F., {Possenti}, A., {Colpi}, M., \& {Spera}, M. 2019, \apjl, 884, L9,
  \dodoi{10.3847/2041-8213/ab46c3}

\bibitem[{{Abbate} {et~al.}(2018){Abbate}, {Possenti}, {Ridolfi}, {Freire},
  {Camilo}, {Manchester}, \& {D'Amico}}]{abbate18}
{Abbate}, F., {Possenti}, A., {Ridolfi}, A., {et~al.} 2018, \mnras, 481, 627,
  \dodoi{10.1093/mnras/sty2298}

\bibitem[{{Alpar} {et~al.}(1982){Alpar}, {Cheng}, {Ruderman}, \&
  {Shaham}}]{alpar82}
{Alpar}, M.~A., {Cheng}, A.~F., {Ruderman}, M.~A., \& {Shaham}, J. 1982, \nat,
  300, 728, \dodoi{10.1038/300728a0}

\bibitem[{{Althaus} {et~al.}(2001{\natexlab{a}}){Althaus}, {Serenelli}, \&
  {Benvenuto}}]{althaus01a}
{Althaus}, L.~G., {Serenelli}, A.~M., \& {Benvenuto}, O.~G. 2001{\natexlab{a}},
  \mnras, 323, 471, \dodoi{10.1046/j.1365-8711.2001.04227.x}

\bibitem[{{Althaus} {et~al.}(2001{\natexlab{b}}){Althaus}, {Serenelli}, \&
  {Benvenuto}}]{althaus01b}
---. 2001{\natexlab{b}}, \mnras, 324, 617,
  \dodoi{10.1046/j.1365-8711.2001.04324.x}

\bibitem[{{Antoniadis} {et~al.}(2016{\natexlab{a}}){Antoniadis}, {Kaplan},
  {Stovall}, {Freire}, {Deneva}, {Koester}, {Jenet}, \&
  {Martinez}}]{antoniadis16wd}
{Antoniadis}, J., {Kaplan}, D.~L., {Stovall}, K., {et~al.} 2016{\natexlab{a}},
  \apj, 830, 36, \dodoi{10.3847/0004-637X/830/1/36}

\bibitem[{{Antoniadis} {et~al.}(2016{\natexlab{b}}){Antoniadis}, {Tauris},
  {Ozel}, {Barr}, {Champion}, \& {Freire}}]{antoniadis16}
{Antoniadis}, J., {Tauris}, T.~M., {Ozel}, F., {et~al.} 2016{\natexlab{b}},
  arXiv e-prints, arXiv:1605.01665.
\newblock \doarXiv{1605.01665}

\bibitem[{{Antoniadis} {et~al.}(2012){Antoniadis}, {van Kerkwijk}, {Koester},
  {Freire}, {Wex}, {Tauris}, {Kramer}, \& {Bassa}}]{antoniadis12}
{Antoniadis}, J., {van Kerkwijk}, M.~H., {Koester}, D., {et~al.} 2012, \mnras,
  423, 3316, \dodoi{10.1111/j.1365-2966.2012.21124.x}

\bibitem[{{Antoniadis} {et~al.}(2013){Antoniadis}, {Freire}, {Wex}, {Tauris},
  {Lynch}, {van Kerkwijk}, {Kramer}, {Bassa}, {Dhillon}, {Driebe}, {Hessels},
  {Kaspi}, {Kondratiev}, {Langer}, {Marsh}, {McLaughlin}, {Pennucci}, {Ransom},
  {Stairs}, {van Leeuwen}, {Verbiest}, \& {Whelan}}]{antoniadis13}
{Antoniadis}, J., {Freire}, P. C.~C., {Wex}, N., {et~al.} 2013, Science, 340,
  448, \dodoi{10.1126/science.1233232}

\bibitem[{{Archibald} {et~al.}(2010){Archibald}, {Kaspi}, {Bogdanov},
  {Hessels}, {Stairs}, {Ransom}, \& {McLaughlin}}]{archibald10}
{Archibald}, A.~M., {Kaspi}, V.~M., {Bogdanov}, S., {et~al.} 2010, \apj, 722,
  88, \dodoi{10.1088/0004-637X/722/1/88}

\bibitem[{{Bellini} {et~al.}(2011){Bellini}, {Anderson}, \&
  {Bedin}}]{bellini11}
{Bellini}, A., {Anderson}, J., \& {Bedin}, L.~R. 2011, \pasp, 123, 622,
  \dodoi{10.1086/659878}

\bibitem[{{Bhattacharya} \& {van den Heuvel}(1991)}]{bhattacharya91}
{Bhattacharya}, D., \& {van den Heuvel}, E.~P.~J. 1991, \physrep, 203, 1,
  \dodoi{10.1016/0370-1573(91)90064-S}

\bibitem[{{Bogdanov} {et~al.}(2019){Bogdanov}, {Guillot}, {Ray}, {Wolff},
  {Chakrabarty}, {Ho}, {Kerr}, {Lamb}, {Lommen}, {Ludlam}, {Milburn},
  {Montano}, {Miller}, {Baub{\"o}ck}, {{\"O}zel}, {Psaltis}, {Remillard},
  {Riley}, {Steiner}, {Strohmayer}, {Watts}, {Wood}, {Zeldes}, {Enoto},
  {Okajima}, {Kellogg}, {Baker}, {Markwardt}, {Arzoumanian}, \&
  {Gendreau}}]{bogdanov19}
{Bogdanov}, S., {Guillot}, S., {Ray}, P.~S., {et~al.} 2019, \apjl, 887, L25,
  \dodoi{10.3847/2041-8213/ab53eb}

\bibitem[{{Bohlin}(2016)}]{bohlin16}
{Bohlin}, R.~C. 2016, \aj, 152, 60, \dodoi{10.3847/0004-6256/152/3/60}

\bibitem[{{Breton} {et~al.}(2013){Breton}, {van Kerkwijk}, {Roberts},
  {Hessels}, {Camilo}, {McLaughlin}, {Ransom}, {Ray}, \& {Stairs}}]{breton13}
{Breton}, R.~P., {van Kerkwijk}, M.~H., {Roberts}, M.~S.~E., {et~al.} 2013,
  \apj, 769, 108, \dodoi{10.1088/0004-637X/769/2/108}

\bibitem[{{Cadelano}(2019)}]{cadelanotesi}
{Cadelano}, M. 2019, arXiv e-prints, arXiv:1901.04212.
\newblock \doarXiv{1901.04212}

\bibitem[{{Cadelano} {et~al.}(2019){Cadelano}, {Ferraro}, {Istrate},
  {Pallanca}, {Lanzoni}, \& {Freire}}]{cadelano19}
{Cadelano}, M., {Ferraro}, F.~R., {Istrate}, A.~G., {et~al.} 2019, \apj, 875,
  25, \dodoi{10.3847/1538-4357/ab0e6b}

\bibitem[{{Cadelano} {et~al.}(2017){Cadelano}, {Pallanca}, {Ferraro},
  {Dalessand ro}, {Lanzoni}, \& {Patruno}}]{cadelano17x}
{Cadelano}, M., {Pallanca}, C., {Ferraro}, F.~R., {et~al.} 2017, \apj, 844, 53,
  \dodoi{10.3847/1538-4357/aa7b7f}

\bibitem[{{Cadelano} {et~al.}(2015{\natexlab{a}}){Cadelano}, {Pallanca},
  {Ferraro}, {Salaris}, {Dalessandro}, {Lanzoni}, \&
  {Freire}}]{cadelano15_47tuc}
---. 2015{\natexlab{a}}, \apj, 812, 63, \dodoi{10.1088/0004-637X/812/1/63}

\bibitem[{{Cadelano} {et~al.}(2018){Cadelano}, {Ransom}, {Freire}, {Ferraro},
  {Hessels}, {Lanzoni}, {Pallanca}, \& {Stairs}}]{cadelano18}
{Cadelano}, M., {Ransom}, S.~M., {Freire}, P.~C.~C., {et~al.} 2018, \apj, 855,
  125, \dodoi{10.3847/1538-4357/aaac2a}

\bibitem[{{Cadelano} {et~al.}(2015{\natexlab{b}}){Cadelano}, {Pallanca},
  {Ferraro}, {Stairs}, {Ransom}, {Dalessandro}, {Lanzoni}, {Hessels}, \&
  {Freire}}]{cadelano15_m71}
{Cadelano}, M., {Pallanca}, C., {Ferraro}, F.~R., {et~al.} 2015{\natexlab{b}},
  \apj, 807, 91, \dodoi{10.1088/0004-637X/807/1/91}

\bibitem[{{Campos} {et~al.}(2018){Campos}, {Pelisoli}, {Kamann}, {Husser},
  {Dreizler}, {Bellini}, {Robinson}, {Nardiello}, {Piotto}, {Kepler},
  {Istrate}, {Winget}, {Montgomery}, \& {Dotter}}]{campos18}
{Campos}, F., {Pelisoli}, I., {Kamann}, S., {et~al.} 2018, \mnras, 481, 4397,
  \dodoi{10.1093/mnras/sty2591}

\bibitem[{{Cardelli} {et~al.}(1989){Cardelli}, {Clayton}, \&
  {Mathis}}]{cardelli89}
{Cardelli}, J.~A., {Clayton}, G.~C., \& {Mathis}, J.~S. 1989, \apj, 345, 245,
  \dodoi{10.1086/167900}

\bibitem[{{Carretta} {et~al.}(2009){Carretta}, {Bragaglia}, {Gratton},
  {D'Orazi}, \& {Lucatello}}]{carretta09}
{Carretta}, E., {Bragaglia}, A., {Gratton}, R., {D'Orazi}, V., \& {Lucatello},
  S. 2009, \aap, 508, 695, \dodoi{10.1051/0004-6361/200913003}

\bibitem[{{Cheng} {et~al.}(2019{\natexlab{a}}){Cheng}, {Li}, {Fang}, {Li}, \&
  {Xu}}]{cheng19}
{Cheng}, Z., {Li}, Z., {Fang}, T., {Li}, X., \& {Xu}, X. 2019{\natexlab{a}},
  \apj, 883, 90, \dodoi{10.3847/1538-4357/ab3c6d}

\bibitem[{{Cheng} {et~al.}(2019{\natexlab{b}}){Cheng}, {Li}, {Li}, {Xu}, \&
  {Fang}}]{cheng1947tuc}
{Cheng}, Z., {Li}, Z., {Li}, X., {Xu}, X., \& {Fang}, T. 2019{\natexlab{b}},
  \apj, 876, 59, \dodoi{10.3847/1538-4357/ab1593}

\bibitem[{{Cocozza} {et~al.}(2008){Cocozza}, {Ferraro}, {Possenti}, {Beccari},
  {Lanzoni}, {Ransom}, {Rood}, \& {D'Amico}}]{cocozza06}
{Cocozza}, G., {Ferraro}, F.~R., {Possenti}, A., {et~al.} 2008, \apjl, 679,
  L105, \dodoi{10.1086/589557}

\bibitem[{{Corongiu} {et~al.}(2012){Corongiu}, {Burgay}, {Possenti}, {Camilo},
  {D'Amico}, {Lyne}, {Manchester}, {Sarkissian}, {Bailes}, {Johnston},
  {Kramer}, \& {van Straten}}]{corongiu12}
{Corongiu}, A., {Burgay}, M., {Possenti}, A., {et~al.} 2012, \apj, 760, 100,
  \dodoi{10.1088/0004-637X/760/2/100}

\bibitem[{{Cromartie} {et~al.}(2020){Cromartie}, {Fonseca}, {Ransom},
  {Demorest}, {Arzoumanian}, {Blumer}, {Brook}, {DeCesar}, {Dolch}, {Ellis},
  {Ferdman}, {Ferrara}, {Garver-Daniels}, {Gentile}, {Jones}, {Lam}, {Lorimer},
  {Lynch}, {McLaughlin}, {Ng}, {Nice}, {Pennucci}, {Spiewak}, {Stairs},
  {Stovall}, {Swiggum}, \& {Zhu}}]{cromartie20}
{Cromartie}, H.~T., {Fonseca}, E., {Ransom}, S.~M., {et~al.} 2020, Nature
  Astronomy, 4, 72, \dodoi{10.1038/s41550-019-0880-2}

\bibitem[{{Dai} {et~al.}(2017){Dai}, {Smith}, {Wang}, {Okamoto}, {Xu}, {Yue},
  \& {Liu}}]{dai17}
{Dai}, S., {Smith}, M.~C., {Wang}, S., {et~al.} 2017, \apj, 842, 105,
  \dodoi{10.3847/1538-4357/aa7209}

\bibitem[{{Demorest} {et~al.}(2010){Demorest}, {Pennucci}, {Ransom}, {Roberts},
  \& {Hessels}}]{demorest10}
{Demorest}, P.~B., {Pennucci}, T., {Ransom}, S.~M., {Roberts}, M.~S.~E., \&
  {Hessels}, J.~W.~T. 2010, \nat, 467, 1081, \dodoi{10.1038/nature09466}

\bibitem[{{Dotter} {et~al.}(2010){Dotter}, {Sarajedini}, {Anderson},
  {Aparicio}, {Bedin}, {Chaboyer}, {Majewski}, {Mar{\'\i}n-Franch}, {Milone},
  {Paust}, {Piotto}, {Reid}, {Rosenberg}, \& {Siegel}}]{dotter10}
{Dotter}, A., {Sarajedini}, A., {Anderson}, J., {et~al.} 2010, \apj, 708, 698,
  \dodoi{10.1088/0004-637X/708/1/698}

\bibitem[{{Driebe} {et~al.}(1998){Driebe}, {Schoenberner}, {Bloecker}, \&
  {Herwig}}]{driebe98}
{Driebe}, T., {Schoenberner}, D., {Bloecker}, T., \& {Herwig}, F. 1998, \aap,
  339, 123.
\newblock \doarXiv{astro-ph/9809079}

\bibitem[{{Edmonds} {et~al.}(2001){Edmonds}, {Gilliland}, {Heinke}, {Grindlay},
  \& {Camilo}}]{edmonds01}
{Edmonds}, P.~D., {Gilliland}, R.~L., {Heinke}, C.~O., {Grindlay}, J.~E., \&
  {Camilo}, F. 2001, \apjl, 557, L57, \dodoi{10.1086/323122}

\bibitem[{{Ferraro} {et~al.}(2019){Ferraro}, {Lanzoni}, {Dalessandro},
  {Cadelano}, {Raso}, {Mucciarelli}, {Beccari}, \& {Pallanca}}]{ferraro19}
{Ferraro}, F.~R., {Lanzoni}, B., {Dalessandro}, E., {et~al.} 2019, Nature
  Astronomy, 3, 1149, \dodoi{10.1038/s41550-019-0865-1}

\bibitem[{{Ferraro} {et~al.}(2016){Ferraro}, {Lapenna}, {Mucciarelli},
  {Lanzoni}, {Dalessandro}, {Pallanca}, \& {Massari}}]{ferraro16ebss}
{Ferraro}, F.~R., {Lapenna}, E., {Mucciarelli}, A., {et~al.} 2016, \apj, 816,
  70, \dodoi{10.3847/0004-637X/816/2/70}

\bibitem[{{Ferraro} {et~al.}(1999){Ferraro}, {Messineo}, {Fusi Pecci}, {de
  Palo}, {Straniero}, {Chieffi}, \& {Limongi}}]{ferraro99}
{Ferraro}, F.~R., {Messineo}, M., {Fusi Pecci}, F., {et~al.} 1999, \aj, 118,
  1738, \dodoi{10.1086/301029}

\bibitem[{{Ferraro} {et~al.}(2015){Ferraro}, {Pallanca}, {Lanzoni}, {Cadelano},
  {Massari}, {Dalessandro}, \& {Mucciarelli}}]{ferraro15}
{Ferraro}, F.~R., {Pallanca}, C., {Lanzoni}, B., {et~al.} 2015, \apjl, 807, L1,
  \dodoi{10.1088/2041-8205/807/1/L1}

\bibitem[{{Ferraro} {et~al.}(1997){Ferraro}, {Paltrinieri}, {Fusi Pecci},
  {Cacciari}, {Dorman}, \& {Rood}}]{ferraro97}
{Ferraro}, F.~R., {Paltrinieri}, B., {Fusi Pecci}, F., {et~al.} 1997, \apjl,
  484, L145, \dodoi{10.1086/310780}

\bibitem[{{Ferraro} {et~al.}(2001){Ferraro}, {Possenti}, {D'Amico}, \&
  {Sabbi}}]{ferraro01}
{Ferraro}, F.~R., {Possenti}, A., {D'Amico}, N., \& {Sabbi}, E. 2001, \apjl,
  561, L93, \dodoi{10.1086/324563}

\bibitem[{{Ferraro} {et~al.}(2003{\natexlab{a}}){Ferraro}, {Possenti}, {Sabbi},
  \& {D'Amico}}]{ferraro03}
{Ferraro}, F.~R., {Possenti}, A., {Sabbi}, E., \& {D'Amico}, N.
  2003{\natexlab{a}}, \apjl, 596, L211, \dodoi{10.1086/379536}

\bibitem[{{Ferraro} {et~al.}(2003{\natexlab{b}}){Ferraro}, {Sabbi}, {Gratton},
  {Possenti}, {D'Amico}, {Bragaglia}, \& {Camilo}}]{ferraro03massratio}
{Ferraro}, F.~R., {Sabbi}, E., {Gratton}, R., {et~al.} 2003{\natexlab{b}},
  \apjl, 584, L13, \dodoi{10.1086/368279}

\bibitem[{{Ferraro} {et~al.}(2009){Ferraro}, {Beccari}, {Dalessandro},
  {Lanzoni}, {Sills}, {Rood}, {Pecci}, {Karakas}, {Miocchi}, \&
  {Bovinelli}}]{ferraro09}
{Ferraro}, F.~R., {Beccari}, G., {Dalessandro}, E., {et~al.} 2009, \nat, 462,
  1028, \dodoi{10.1038/nature08607}

\bibitem[{{Ferraro} {et~al.}(2012){Ferraro}, {Lanzoni}, {Dalessandro},
  {Beccari}, {Pasquato}, {Miocchi}, {Rood}, {Sigurdsson}, {Sills}, {Vesperini},
  {Mapelli}, {Contreras}, {Sanna}, \& {Mucciarelli}}]{ferraro12}
{Ferraro}, F.~R., {Lanzoni}, B., {Dalessandro}, E., {et~al.} 2012, \nat, 492,
  393, \dodoi{10.1038/nature11686}

\bibitem[{{Ferraro} {et~al.}(2018){Ferraro}, {Lanzoni}, {Raso}, {Nardiello},
  {Dalessandro}, {Vesperini}, {Piotto}, {Pallanca}, {Beccari}, {Bellini},
  {Libralato}, {Anderson}, {Aparicio}, {Bedin}, {Cassisi}, {Milone},
  {Ortolani}, {Renzini}, {Salaris}, \& {van der Marel}}]{ferraro18}
{Ferraro}, F.~R., {Lanzoni}, B., {Raso}, S., {et~al.} 2018, \apj, 860, 36,
  \dodoi{10.3847/1538-4357/aac01c}

\bibitem[{{Fonseca} {et~al.}(2016){Fonseca}, {Pennucci}, {Ellis}, {Stairs},
  {Nice}, {Ransom}, {Demorest}, {Arzoumanian}, {Crowter}, {Dolch}, {Ferdman},
  {Gonzalez}, {Jones}, {Jones}, {Lam}, {Levin}, {McLaughlin}, {Stovall},
  {Swiggum}, \& {Zhu}}]{fonseca16}
{Fonseca}, E., {Pennucci}, T.~T., {Ellis}, J.~A., {et~al.} 2016, \apj, 832,
  167, \dodoi{10.3847/0004-637X/832/2/167}

\bibitem[{{Foreman-Mackey}(2016)}]{foreman16}
{Foreman-Mackey}, D. 2016, The Journal of Open Source Software, 1, 24,
  \dodoi{10.21105/joss.00024}

\bibitem[{{Foreman-Mackey} {et~al.}(2019{\natexlab{a}}){Foreman-Mackey},
  {Farr}, {Sinha}, {Archibald}, {Hogg}, {Sanders}, {Zuntz}, {Williams},
  {Nelson}, {de Val-Borro}, {Erhardt}, {Pashchenko}, \& {Pla}}]{emcee_v3}
{Foreman-Mackey}, D., {Farr}, W., {Sinha}, M., {et~al.} 2019{\natexlab{a}}, The
  Journal of Open Source Software, 4, 1864, \dodoi{10.21105/joss.01864}

\bibitem[{{Foreman-Mackey} {et~al.}(2019{\natexlab{b}}){Foreman-Mackey},
  {Farr}, {Sinha}, {Archibald}, {Hogg}, {Sanders}, {Zuntz}, {Williams},
  {Nelson}, {de Val-Borro}, {Erhardt}, {Pashchenko}, \& {Pla}}]{foreman19}
---. 2019{\natexlab{b}}, The Journal of Open Source Software, 4, 1864,
  \dodoi{10.21105/joss.01864}

\bibitem[{{Freire} \& {Wex}(2010)}]{freire10}
{Freire}, P. C.~C., \& {Wex}, N. 2010, \mnras, 409, 199,
  \dodoi{10.1111/j.1365-2966.2010.17319.x}

\bibitem[{{Freire} {et~al.}(2017){Freire}, {Ridolfi}, {Kramer}, {Jordan},
  {Manchester}, {Torne}, {Sarkissian}, {Heinke}, {D'Amico}, {Camilo},
  {Lorimer}, \& {Lyne}}]{freire17}
{Freire}, P.~C.~C., {Ridolfi}, A., {Kramer}, M., {et~al.} 2017, \mnras, 471,
  857, \dodoi{10.1093/mnras/stx1533}

\bibitem[{{Gaia Collaboration} {et~al.}(2018){Gaia Collaboration}, {Brown},
  {Vallenari}, {Prusti}, {de Bruijne}, {Babusiaux}, {Bailer-Jones}, {Biermann},
  {Evans}, {Eyer}, {Jansen}, {Jordi}, {Klioner}, {Lammers}, {Lindegren},
  {Luri}, {Mignard}, {Panem}, {Pourbaix}, {Randich}, {Sartoretti}, {Siddiqui},
  {Soubiran}, {van Leeuwen}, {Walton}, {Arenou}, {Bastian}, {Cropper},
  {Drimmel}, {Katz}, {Lattanzi}, {Bakker}, {Cacciari}, {Casta{\~n}eda},
  {Chaoul}, {Cheek}, {De Angeli}, {Fabricius}, {Guerra}, {Holl}, {Masana},
  {Messineo}, {Mowlavi}, {Nienartowicz}, {Panuzzo}, {Portell}, {Riello},
  {Seabroke}, {Tanga}, {Th{\'e}venin}, {Gracia-Abril}, {Comoretto},
  {Garcia-Reinaldos}, {Teyssier}, {Altmann}, {Andrae}, {Audard},
  {Bellas-Velidis}, {Benson}, {Berthier}, {Blomme}, {Burgess}, {Busso},
  {Carry}, {Cellino}, {Clementini}, {Clotet}, {Creevey}, {Davidson}, {De
  Ridder}, {Delchambre}, {Dell'Oro}, {Ducourant},
  {Fern{\'a}ndez-Hern{\'a}ndez}, {Fouesneau}, {Fr{\'e}mat}, {Galluccio},
  {Garc{\'\i}a-Torres}, {Gonz{\'a}lez-N{\'u}{\~n}ez}, {Gonz{\'a}lez-Vidal},
  {Gosset}, {Guy}, {Halbwachs}, {Hambly}, {Harrison}, {Hern{\'a}ndez},
  {Hestroffer}, {Hodgkin}, {Hutton}, {Jasniewicz}, {Jean-Antoine-Piccolo},
  {Jordan}, {Korn}, {Krone-Martins}, {Lanzafame}, {Lebzelter}, {L{\"o}ffler},
  {Manteiga}, {Marrese}, {Mart{\'\i}n-Fleitas}, {Moitinho}, {Mora}, {Muinonen},
  {Osinde}, {Pancino}, {Pauwels}, {Petit}, {Recio-Blanco}, {Richards},
  {Rimoldini}, {Robin}, {Sarro}, {Siopis}, {Smith}, {Sozzetti}, {S{\"u}veges},
  {Torra}, {van Reeven}, {Abbas}, {Abreu Aramburu}, {Accart}, {Aerts},
  {Altavilla}, {{\'A}lvarez}, {Alvarez}, {Alves}, {Anderson}, {Andrei},
  {Anglada Varela}, {Antiche}, {Antoja}, {Arcay}, {Astraatmadja}, {Bach},
  {Baker}, {Balaguer-N{\'u}{\~n}ez}, {Balm}, {Barache}, {Barata}, {Barbato},
  {Barblan}, {Barklem}, {Barrado}, {Barros}, {Barstow}, {Bartholom{\'e}
  Mu{\~n}oz}, {Bassilana}, {Becciani}, {Bellazzini}, {Berihuete}, {Bertone},
  {Bianchi}, {Bienaym{\'e}}, {Blanco-Cuaresma}, {Boch}, {Boeche}, {Bombrun},
  {Borrachero}, {Bossini}, {Bouquillon}, {Bourda}, {Bragaglia}, {Bramante},
  {Breddels}, {Bressan}, {Brouillet}, {Br{\"u}semeister}, {Brugaletta},
  {Bucciarelli}, {Burlacu}, {Busonero}, {Butkevich}, {Buzzi}, {Caffau},
  {Cancelliere}, {Cannizzaro}, {Cantat-Gaudin}, {Carballo}, {Carlucci},
  {Carrasco}, {Casamiquela}, {Castellani}, {Castro-Ginard}, {Charlot},
  {Chemin}, {Chiavassa}, {Cocozza}, {Costigan}, {Cowell}, {Crifo}, {Crosta},
  {Crowley}, {Cuypers}, {Dafonte}, {Damerdji}, {Dapergolas}, {David}, {David},
  {de Laverny}, {De Luise}, {De March}, {de Martino}, {de Souza}, {de Torres},
  {Debosscher}, {del Pozo}, {Delbo}, {Delgado}, {Delgado}, {Di Matteo},
  {Diakite}, {Diener}, {Distefano}, {Dolding}, {Drazinos}, {Dur{\'a}n},
  {Edvardsson}, {Enke}, {Eriksson}, {Esquej}, {Eynard Bontemps}, {Fabre},
  {Fabrizio}, {Faigler}, {Falc{\~a}o}, {Farr{\`a}s Casas}, {Federici},
  {Fedorets}, {Fernique}, {Figueras}, {Filippi}, {Findeisen}, {Fonti},
  {Fraile}, {Fraser}, {Fr{\'e}zouls}, {Gai}, {Galleti}, {Garabato},
  {Garc{\'\i}a-Sedano}, {Garofalo}, {Garralda}, {Gavel}, {Gavras}, {Gerssen},
  {Geyer}, {Giacobbe}, {Gilmore}, {Girona}, {Giuffrida}, {Glass}, {Gomes},
  {Granvik}, {Gueguen}, {Guerrier}, {Guiraud}, {Guti{\'e}rrez-S{\'a}nchez},
  {Haigron}, {Hatzidimitriou}, {Hauser}, {Haywood}, {Heiter}, {Helmi}, {Heu},
  {Hilger}, {Hobbs}, {Hofmann}, {Holland}, {Huckle}, {Hypki}, {Icardi},
  {Jan{\ss}en}, {Jevardat de Fombelle}, {Jonker}, {Juh{\'a}sz}, {Julbe},
  {Karampelas}, {Kewley}, {Klar}, {Kochoska}, {Kohley}, {Kolenberg},
  {Kontizas}, {Kontizas}, {Koposov}, {Kordopatis}, {Kostrzewa-Rutkowska},
  {Koubsky}, {Lambert}, {Lanza}, {Lasne}, {Lavigne}, {Le Fustec}, {Le
  Poncin-Lafitte}, {Lebreton}, {Leccia}, {Leclerc}, {Lecoeur-Taibi},
  {Lenhardt}, {Leroux}, {Liao}, {Licata}, {Lindstr{\o}m}, {Lister}, {Livanou},
  {Lobel}, {L{\'o}pez}, {Managau}, {Mann}, {Mantelet}, {Marchal}, {Marchant},
  {Marconi}, {Marinoni}, {Marschalk{\'o}}, {Marshall}, {Martino}, {Marton},
  {Mary}, {Massari}, {Matijevi{\v{c}}}, {Mazeh}, {McMillan}, {Messina},
  {Michalik}, {Millar}, {Molina}, {Molinaro}, {Moln{\'a}r}, {Montegriffo},
  {Mor}, {Morbidelli}, {Morel}, {Morris}, {Mulone}, {Muraveva}, {Musella},
  {Nelemans}, {Nicastro}, {Noval}, {O'Mullane}, {Ord{\'e}novic},
  {Ord{\'o}{\~n}ez-Blanco}, {Osborne}, {Pagani}, {Pagano}, {Pailler},
  {Palacin}, {Palaversa}, {Panahi}, {Pawlak}, {Piersimoni}, {Pineau}, {Plachy},
  {Plum}, {Poggio}, {Poujoulet}, {Pr{\v{s}}a}, {Pulone}, {Racero}, {Ragaini},
  {Rambaux}, {Ramos-Lerate}, {Regibo}, {Reyl{\'e}}, {Riclet}, {Ripepi}, {Riva},
  {Rivard}, {Rixon}, {Roegiers}, {Roelens}, {Romero-G{\'o}mez}, {Rowell},
  {Royer}, {Ruiz-Dern}, {Sadowski}, {Sagrist{\`a} Sell{\'e}s}, {Sahlmann},
  {Salgado}, {Salguero}, {Sanna}, {Santana-Ros}, {Sarasso}, {Savietto},
  {Schultheis}, {Sciacca}, {Segol}, {Segovia}, {S{\'e}gransan}, {Shih},
  {Siltala}, {Silva}, {Smart}, {Smith}, {Solano}, {Solitro}, {Sordo}, {Soria
  Nieto}, {Souchay}, {Spagna}, {Spoto}, {Stampa}, {Steele},
  {Steidelm{\"u}ller}, {Stephenson}, {Stoev}, {Suess}, {Surdej}, {Szabados},
  {Szegedi-Elek}, {Tapiador}, {Taris}, {Tauran}, {Taylor}, {Teixeira},
  {Terrett}, {Teyssand ier}, {Thuillot}, {Titarenko}, {Torra Clotet}, {Turon},
  {Ulla}, {Utrilla}, {Uzzi}, {Vaillant}, {Valentini}, {Valette}, {van Elteren},
  {Van Hemelryck}, {van Leeuwen}, {Vaschetto}, {Vecchiato}, {Veljanoski},
  {Viala}, {Vicente}, {Vogt}, {von Essen}, {Voss}, {Votruba}, {Voutsinas},
  {Walmsley}, {Weiler}, {Wertz}, {Wevers}, {Wyrzykowski}, {Yoldas},
  {{\v{Z}}erjal}, {Ziaeepour}, {Zorec}, {Zschocke}, {Zucker}, {Zurbach}, \&
  {Zwitter}}]{gaia18}
{Gaia Collaboration}, {Brown}, A.~G.~A., {Vallenari}, A., {et~al.} 2018, \aap,
  616, A1, \dodoi{10.1051/0004-6361/201833051}

\bibitem[{{Harris}(1996)}]{harris96}
{Harris}, W.~E. 1996, \aj, 112, 1487, \dodoi{10.1086/118116}

\bibitem[{{Hessels} {et~al.}(2007){Hessels}, {Ransom}, {Stairs}, {Kaspi}, \&
  {Freire}}]{hessels07}
{Hessels}, J.~W.~T., {Ransom}, S.~M., {Stairs}, I.~H., {Kaspi}, V.~M., \&
  {Freire}, P.~C.~C. 2007, \apj, 670, 363, \dodoi{10.1086/521780}

\bibitem[{{Hong} {et~al.}(2017){Hong}, {Vesperini}, {Belloni}, \&
  {Giersz}}]{hong17}
{Hong}, J., {Vesperini}, E., {Belloni}, D., \& {Giersz}, M. 2017, \mnras, 464,
  2511, \dodoi{10.1093/mnras/stw2595}

\bibitem[{{Istrate} {et~al.}(2016){Istrate}, {Marchant}, {Tauris}, {Langer},
  {Stancliffe}, \& {Grassitelli}}]{istrate16}
{Istrate}, A.~G., {Marchant}, P., {Tauris}, T.~M., {et~al.} 2016, \aap, 595,
  A35, \dodoi{10.1051/0004-6361/201628874}

\bibitem[{{Istrate} {et~al.}(2014{\natexlab{a}}){Istrate}, {Tauris}, \&
  {Langer}}]{istrate14a}
{Istrate}, A.~G., {Tauris}, T.~M., \& {Langer}, N. 2014{\natexlab{a}}, \aap,
  571, A45, \dodoi{10.1051/0004-6361/201424680}

\bibitem[{{Istrate} {et~al.}(2014{\natexlab{b}}){Istrate}, {Tauris}, {Langer},
  \& {Antoniadis}}]{istrate14b}
{Istrate}, A.~G., {Tauris}, T.~M., {Langer}, N., \& {Antoniadis}, J.
  2014{\natexlab{b}}, \aap, 571, L3, \dodoi{10.1051/0004-6361/201424681}

\bibitem[{{Jacoby} {et~al.}(2006){Jacoby}, {Cameron}, {Jenet}, {Anderson},
  {Murty}, \& {Kulkarni}}]{jacoby06}
{Jacoby}, B.~A., {Cameron}, P.~B., {Jenet}, F.~A., {et~al.} 2006, \apjl, 644,
  L113, \dodoi{10.1086/505742}

\bibitem[{{Joss} {et~al.}(1987){Joss}, {Rappaport}, \& {Lewis}}]{joss1987}
{Joss}, P.~C., {Rappaport}, S., \& {Lewis}, W. 1987, \apj, 319, 180,
  \dodoi{10.1086/165443}

\bibitem[{{Kaplan} {et~al.}(2012){Kaplan}, {Stovall}, {Ransom}, {Roberts},
  {Kotulla}, {Archibald}, {Biwer}, {Boyles}, {Dartez}, {Day}, {Ford}, {Garcia},
  {Hessels}, {Jenet}, {Karako}, {Kaspi}, {Kondratiev}, {Lorimer}, {Lynch},
  {McLaughlin}, {Rohr}, {Siemens}, {Stairs}, \& {van Leeuwen}}]{kaplan12}
{Kaplan}, D.~L., {Stovall}, K., {Ransom}, S.~M., {et~al.} 2012, \apj, 753, 174,
  \dodoi{10.1088/0004-637X/753/2/174}

\bibitem[{{Kilic} {et~al.}(2015){Kilic}, {Hermes}, {Gianninas}, \&
  {Brown}}]{kilic15}
{Kilic}, M., {Hermes}, J.~J., {Gianninas}, A., \& {Brown}, W.~R. 2015, \mnras,
  446, L26, \dodoi{10.1093/mnrasl/slu152}

\bibitem[{{Kirichenko} {et~al.}(2020){Kirichenko}, {Karpova}, {Zyuzin},
  {Zharikov}, {L{\'o}pez}, {Shibanov}, {Freire}, {Fonseca}, \&
  {Cabrera-Lavers}}]{kirichenko20}
{Kirichenko}, A.~Y., {Karpova}, A.~V., {Zyuzin}, D.~A., {et~al.} 2020, \mnras,
  492, 3032, \dodoi{10.1093/mnras/staa066}

\bibitem[{{Koester}(2010)}]{koester10}
{Koester}, D. 2010, \memsai, 81, 921

\bibitem[{{Kulkarni} {et~al.}(1991){Kulkarni}, {Anderson}, {Prince}, \&
  {Wolszczan}}]{kulkarni91}
{Kulkarni}, S.~R., {Anderson}, S.~B., {Prince}, T.~A., \& {Wolszczan}, A. 1991,
  \nat, 349, 47, \dodoi{10.1038/349047a0}

\bibitem[{{Lattimer} \& {Prakash}(2001)}]{lattimer01}
{Lattimer}, J.~M., \& {Prakash}, M. 2001, \apj, 550, 426,
  \dodoi{10.1086/319702}

\bibitem[{{Lattimer} \& {Prakash}(2007)}]{lattimer07}
---. 2007, \physrep, 442, 109, \dodoi{10.1016/j.physrep.2007.02.003}

\bibitem[{{Lin} {et~al.}(2011){Lin}, {Rappaport}, {Podsiadlowski}, {Nelson},
  {Paxton}, \& {Todorov}}]{lin2011}
{Lin}, J., {Rappaport}, S., {Podsiadlowski}, P., {et~al.} 2011, \apj, 732, 70,
  \dodoi{10.1088/0004-637X/732/2/70}

\bibitem[{{Lorimer} \& {Kramer}(2012)}]{handbook}
{Lorimer}, D.~R., \& {Kramer}, M. 2012, {Handbook of Pulsar Astronomy}

\bibitem[{{Mata S{\'a}nchez} {et~al.}(2020){Mata S{\'a}nchez}, {Istrate}, {van
  Kerkwijk}, {Breton}, \& {Kaplan}}]{mata20}
{Mata S{\'a}nchez}, D., {Istrate}, A.~G., {van Kerkwijk}, M.~H., {Breton},
  R.~P., \& {Kaplan}, D.~L. 2020, arXiv e-prints, arXiv:2004.02901.
\newblock \doarXiv{2004.02901}

\bibitem[{{Maxted} {et~al.}(2013){Maxted}, {Serenelli}, {Miglio}, {Marsh},
  {Heber}, {Dhillon}, {Littlefair}, {Copperwheat}, {Smalley}, {Breedt}, \&
  {Schaffenroth}}]{maxted13}
{Maxted}, P. F.~L., {Serenelli}, A.~M., {Miglio}, A., {et~al.} 2013, \nat, 498,
  463, \dodoi{10.1038/nature12192}

\bibitem[{{Meurer} {et~al.}(2003){Meurer}, {Lindler}, {Blakeslee}, {Cox},
  {Martel}, {Tran}, {Bouwens}, {Ford}, {Clampin}, {Hartig}, {Sirianni}, \& {De
  Marchi}}]{meurer03}
{Meurer}, G.~R., {Lindler}, D.~J., {Blakeslee}, J., {et~al.} 2003, Society of
  Photo-Optical Instrumentation Engineers (SPIE) Conference Series, Vol. 4854,
  {Calibration of geometric distortion in the ACS detectors}, ed. J.~C.
  {Blades} \& O.~H.~W. {Siegmund}, 507--514, \dodoi{10.1117/12.460259}

\bibitem[{{Mucciarelli} {et~al.}(2013){Mucciarelli}, {Salaris}, {Lanzoni},
  {Pallanca}, {Dalessandro}, \& {Ferraro}}]{mucciarelli13}
{Mucciarelli}, A., {Salaris}, M., {Lanzoni}, B., {et~al.} 2013, \apjl, 772,
  L27, \dodoi{10.1088/2041-8205/772/2/L27}

\bibitem[{{O'Donnell}(1994)}]{odonnell94}
{O'Donnell}, J.~E. 1994, \apj, 422, 158, \dodoi{10.1086/173713}

\bibitem[{{{\"O}zel} \& {Freire}(2016)}]{ozel16}
{{\"O}zel}, F., \& {Freire}, P. 2016, \araa, 54, 401,
  \dodoi{10.1146/annurev-astro-081915-023322}

\bibitem[{{Pallanca} {et~al.}(2013{\natexlab{a}}){Pallanca}, {Dalessandro},
  {Ferraro}, {Lanzoni}, \& {Beccari}}]{pallanca13}
{Pallanca}, C., {Dalessandro}, E., {Ferraro}, F.~R., {Lanzoni}, B., \&
  {Beccari}, G. 2013{\natexlab{a}}, \apj, 773, 122,
  \dodoi{10.1088/0004-637X/773/2/122}

\bibitem[{{Pallanca} {et~al.}(2013{\natexlab{b}}){Pallanca}, {Lanzoni},
  {Dalessandro}, {Ferraro}, {Possenti}, {Salaris}, \& {Burgay}}]{pallanca13wd}
{Pallanca}, C., {Lanzoni}, B., {Dalessandro}, E., {et~al.} 2013{\natexlab{b}},
  \apj, 773, 127, \dodoi{10.1088/0004-637X/773/2/127}

\bibitem[{{Pallanca} {et~al.}(2014){Pallanca}, {Ransom}, {Ferraro}, {Dalessand
  ro}, {Lanzoni}, {Hessels}, {Stairs}, \& {Freire}}]{pallanca14}
{Pallanca}, C., {Ransom}, S.~M., {Ferraro}, F.~R., {et~al.} 2014, \apj, 795,
  29, \dodoi{10.1088/0004-637X/795/1/29}

\bibitem[{{Pallanca} {et~al.}(2010){Pallanca}, {Dalessandro}, {Ferraro},
  {Lanzoni}, {Rood}, {Possenti}, {D'Amico}, {Freire}, {Stairs}, {Ransom}, \&
  {B{\'e}gin}}]{pallanca10}
{Pallanca}, C., {Dalessandro}, E., {Ferraro}, F.~R., {et~al.} 2010, \apj, 725,
  1165, \dodoi{10.1088/0004-637X/725/1/1165}

\bibitem[{{Papitto} {et~al.}(2013){Papitto}, {Ferrigno}, {Bozzo}, {Rea},
  {Pavan}, {Burderi}, {Burgay}, {Campana}, {di Salvo}, {Falanga},
  {Filipovi{\'c}}, {Freire}, {Hessels}, {Possenti}, {Ransom}, {Riggio},
  {Romano}, {Sarkissian}, {Stairs}, {Stella}, {Torres}, {Wieringa}, \&
  {Wong}}]{papitto13}
{Papitto}, A., {Ferrigno}, C., {Bozzo}, E., {et~al.} 2013, \nat, 501, 517,
  \dodoi{10.1038/nature12470}

\bibitem[{{Parsons} {et~al.}(2020){Parsons}, {Brown}, {Littlefair}, {Dhillon},
  {Marsh}, {Hermes}, {Istrate}, {Breedt}, {Dyer}, {Green}, \&
  {Sahman}}]{parson20}
{Parsons}, S.~G., {Brown}, A.~J., {Littlefair}, S.~P., {et~al.} 2020, arXiv
  e-prints, arXiv:2003.07371.
\newblock \doarXiv{2003.07371}

\bibitem[{{Paxton} {et~al.}(2011){Paxton}, {Bildsten}, {Dotter}, {Herwig},
  {Lesaffre}, \& {Timmes}}]{paxton2011}
{Paxton}, B., {Bildsten}, L., {Dotter}, A., {et~al.} 2011, \apjs, 192, 3,
  \dodoi{10.1088/0067-0049/192/1/3}

\bibitem[{{Paxton} {et~al.}(2013){Paxton}, {Cantiello}, {Arras}, {Bildsten},
  {Brown}, {Dotter}, {Mankovich}, {Montgomery}, {Stello}, {Timmes}, \&
  {Townsend}}]{paxton2013}
{Paxton}, B., {Cantiello}, M., {Arras}, P., {et~al.} 2013, \apjs, 208, 4,
  \dodoi{10.1088/0067-0049/208/1/4}

\bibitem[{{Paxton} {et~al.}(2015){Paxton}, {Marchant}, {Schwab}, {Bauer},
  {Bildsten}, {Cantiello}, {Dessart}, {Farmer}, {Hu}, {Langer}, {Townsend},
  {Townsley}, \& {Timmes}}]{paxton2015}
{Paxton}, B., {Marchant}, P., {Schwab}, J., {et~al.} 2015, \apjs, 220, 15,
  \dodoi{10.1088/0067-0049/220/1/15}

\bibitem[{{Paxton} {et~al.}(2018){Paxton}, {Schwab}, {Bauer}, {Bildsten},
  {Blinnikov}, {Duffell}, {Farmer}, {Goldberg}, {Marchant}, {Sorokina},
  {Thoul}, {Townsend}, \& {Timmes}}]{paxton2018}
{Paxton}, B., {Schwab}, J., {Bauer}, E.~B., {et~al.} 2018, \apjs, 234, 34,
  \dodoi{10.3847/1538-4365/aaa5a8}

\bibitem[{{Pietrinferni} {et~al.}(2004){Pietrinferni}, {Cassisi}, {Salaris}, \&
  {Castelli}}]{pietrinferni04}
{Pietrinferni}, A., {Cassisi}, S., {Salaris}, M., \& {Castelli}, F. 2004, \apj,
  612, 168, \dodoi{10.1086/422498}

\bibitem[{{Pietrinferni} {et~al.}(2006){Pietrinferni}, {Cassisi}, {Salaris}, \&
  {Castelli}}]{pietrinferni06}
---. 2006, \apj, 642, 797, \dodoi{10.1086/501344}

\bibitem[{{Piotto} {et~al.}(2015){Piotto}, {Milone}, {Bedin}, {Anderson},
  {King}, {Marino}, {Nardiello}, {Aparicio}, {Barbuy}, {Bellini}, {Brown},
  {Cassisi}, {Cool}, {Cunial}, {Dalessandro}, {D'Antona}, {Ferraro}, {Hidalgo},
  {Lanzoni}, {Monelli}, {Ortolani}, {Renzini}, {Salaris}, {Sarajedini}, {van
  der Marel}, {Vesperini}, \& {Zoccali}}]{piotto15}
{Piotto}, G., {Milone}, A.~P., {Bedin}, L.~R., {et~al.} 2015, \aj, 149, 91,
  \dodoi{10.1088/0004-6256/149/3/91}

\bibitem[{{Prager} {et~al.}(2017){Prager}, {Ransom}, {Freire}, {Hessels},
  {Stairs}, {Arras}, \& {Cadelano}}]{prager17}
{Prager}, B.~J., {Ransom}, S.~M., {Freire}, P. C.~C., {et~al.} 2017, \apj, 845,
  148, \dodoi{10.3847/1538-4357/aa7ed7}

\bibitem[{{Rappaport} {et~al.}(1995){Rappaport}, {Podsiadlowski}, {Joss}, {Di
  Stefano}, \& {Han}}]{rappaport1995}
{Rappaport}, S., {Podsiadlowski}, P., {Joss}, P.~C., {Di Stefano}, R., \&
  {Han}, Z. 1995, \mnras, 273, 731, \dodoi{10.1093/mnras/273.3.731}

\bibitem[{{Raso} {et~al.}(2017){Raso}, {Ferraro}, {Dalessandro}, {Lanzoni},
  {Nardiello}, {Bellini}, \& {Vesperini}}]{raso17}
{Raso}, S., {Ferraro}, F.~R., {Dalessandro}, E., {et~al.} 2017, \apj, 839, 64,
  \dodoi{10.3847/1538-4357/aa6891}

\bibitem[{{Ridolfi} {et~al.}(2019){Ridolfi}, {Freire}, {Gupta}, \&
  {Ransom}}]{ridolfi19}
{Ridolfi}, A., {Freire}, P.~C.~C., {Gupta}, Y., \& {Ransom}, S.~M. 2019,
  \mnras, 490, 3860, \dodoi{10.1093/mnras/stz2645}

\bibitem[{{Rivera Sandoval} {et~al.}(2018){Rivera Sandoval}, {van den Berg},
  {Heinke}, {Cohn}, {Lugger}, {Anderson}, {Cool}, {Edmonds}, {Wijnands},
  {Ivanova}, \& {Grindlay}}]{riverasandoval18}
{Rivera Sandoval}, L.~E., {van den Berg}, M., {Heinke}, C.~O., {et~al.} 2018,
  \mnras, 475, 4841, \dodoi{10.1093/mnras/sty058}

\bibitem[{{Roberts}(2013)}]{roberts13}
{Roberts}, M. S.~E. 2013, in IAU Symposium, Vol. 291, Neutron Stars and
  Pulsars: Challenges and Opportunities after 80 years, ed. J.~{van Leeuwen},
  127--132, \dodoi{10.1017/S174392131202337X}

\bibitem[{{Roberts} {et~al.}(2018){Roberts}, {Al Noori}, {Torres},
  {McLaughlin}, {Gentile}, {Hessels}, {Ransom}, {Ray}, {Kerr}, \&
  {Breton}}]{roberts18}
{Roberts}, M. S.~E., {Al Noori}, H., {Torres}, R.~A., {et~al.} 2018, in IAU
  Symposium, Vol. 337, Pulsar Astrophysics the Next Fifty Years, ed.
  P.~{Weltevrede}, B.~B.~P. {Perera}, L.~L. {Preston}, \& S.~{Sanidas}, 43--46,
  \dodoi{10.1017/S1743921318000480}

\bibitem[{{Salaris} {et~al.}(2010){Salaris}, {Cassisi}, {Pietrinferni},
  {Kowalski}, \& {Isern}}]{salaris10}
{Salaris}, M., {Cassisi}, S., {Pietrinferni}, A., {Kowalski}, P.~M., \&
  {Isern}, J. 2010, \apj, 716, 1241, \dodoi{10.1088/0004-637X/716/2/1241}

\bibitem[{{Sarajedini} {et~al.}(2007){Sarajedini}, {Bedin}, {Chaboyer},
  {Dotter}, {Siegel}, {Anderson}, {Aparicio}, {King}, {Majewski},
  {Mar{\'\i}n-Franch}, {Piotto}, {Reid}, \& {Rosenberg}}]{sarajedini07}
{Sarajedini}, A., {Bedin}, L.~R., {Chaboyer}, B., {et~al.} 2007, \aj, 133,
  1658, \dodoi{10.1086/511979}

\bibitem[{Savonije(1987)}]{sav87}
Savonije, G.~J. 1987, 325, 416

\bibitem[{{Stappers} {et~al.}(2014){Stappers}, {Archibald}, {Hessels}, {Bassa},
  {Bogdanov}, {Janssen}, {Kaspi}, {Lyne}, {Patruno}, {Tendulkar}, {Hill}, \&
  {Glanzman}}]{stappers14}
{Stappers}, B.~W., {Archibald}, A.~M., {Hessels}, J.~W.~T., {et~al.} 2014,
  \apj, 790, 39, \dodoi{10.1088/0004-637X/790/1/39}

\bibitem[{{Steiner} {et~al.}(2010){Steiner}, {Lattimer}, \&
  {Brown}}]{steiner10}
{Steiner}, A.~W., {Lattimer}, J.~M., \& {Brown}, E.~F. 2010, \apj, 722, 33,
  \dodoi{10.1088/0004-637X/722/1/33}

\bibitem[{{Stetson}(1987)}]{stetson87}
{Stetson}, P.~B. 1987, \pasp, 99, 191, \dodoi{10.1086/131977}

\bibitem[{{Stetson}(1994)}]{stetson94}
---. 1994, \pasp, 106, 250, \dodoi{10.1086/133378}

\bibitem[{{STScI Development Team}(2013)}]{pysynphot}
{STScI Development Team}. 2013, {pysynphot: Synthetic photometry software
  package}.
\newblock \doeprint{1303.023}

\bibitem[{{Tassoul} {et~al.}(1990){Tassoul}, {Fontaine}, \&
  {Winget}}]{tassoul90}
{Tassoul}, M., {Fontaine}, G., \& {Winget}, D.~E. 1990, \apjs, 72, 335,
  \dodoi{10.1086/191420}

\bibitem[{{Tauris}(2012)}]{tauris12a}
{Tauris}, T.~M. 2012, Science, 335, 561, \dodoi{10.1126/science.1216355}

\bibitem[{{Tauris} {et~al.}(2011){Tauris}, {Langer}, \& {Kramer}}]{tauris11}
{Tauris}, T.~M., {Langer}, N., \& {Kramer}, M. 2011, \mnras, 416, 2130,
  \dodoi{10.1111/j.1365-2966.2011.19189.x}

\bibitem[{{Tauris} {et~al.}(2012){Tauris}, {Langer}, \& {Kramer}}]{tauris12b}
---. 2012, \mnras, 425, 1601, \dodoi{10.1111/j.1365-2966.2012.21446.x}

\bibitem[{{Tauris} \& {Savonije}(1999)}]{tauris99}
{Tauris}, T.~M., \& {Savonije}, G.~J. 1999, \aap, 350, 928.
\newblock \doarXiv{astro-ph/9909147}

\bibitem[{{Tauris} \& {van den Heuvel}(2006)}]{tauris06}
{Tauris}, T.~M., \& {van den Heuvel}, E.~P.~J. 2006, {Formation and evolution
  of compact stellar X-ray sources}, Vol.~39, 623--665

\bibitem[{{Tauris} {et~al.}(2017){Tauris}, {Kramer}, {Freire}, {Wex}, {Janka},
  {Langer}, {Podsiadlowski}, {Bozzo}, {Chaty}, {Kruckow}, {van den Heuvel},
  {Antoniadis}, {Breton}, \& {Champion}}]{tauris17}
{Tauris}, T.~M., {Kramer}, M., {Freire}, P.~C.~C., {et~al.} 2017, \apj, 846,
  170, \dodoi{10.3847/1538-4357/aa7e89}

\bibitem[{{Tremblay} \& {Bergeron}(2009)}]{tremblay09}
{Tremblay}, P.~E., \& {Bergeron}, P. 2009, \apj, 696, 1755,
  \dodoi{10.1088/0004-637X/696/2/1755}

\bibitem[{{Wang} {et~al.}(2020){Wang}, {Peng}, {Stappers}, {Liu}, {Keith},
  {Lyne}, {Lu}, {Yu}, {Kou}, {Yan}, {Jiang}, {Jin}, {Li}, {Li}, {Qian}, {Wang},
  {Yue}, {Zhang}, {Zhang}, {Zhu}, \& {FAST Collaboration}}]{wang20}
{Wang}, L., {Peng}, B., {Stappers}, B.~W., {et~al.} 2020, \apj, 892, 43,
  \dodoi{10.3847/1538-4357/ab76cc}

\end{thebibliography}
\bibliographystyle{aasjournal}

\end{document}